\documentclass[12pt]{iopart}

\usepackage{epsfig}

\def\be{\begin{equation}}
\def\ee{\end{equation}}
\def\ba{\begin{eqnarray}}
\def\ea{\end{eqnarray}}

\def\lsim{\raise0.3ex\hbox{$\;<$\kern-0.75em\raise-1.1ex\hbox{$\sim\;$}}}
\def\gsim{\raise0.3ex\hbox{$\;>$\kern-0.75em\raise-1.1ex\hbox{$\sim\;$}}}

\def\dmsq{\Delta m^2}

\def\nue{{\nu_e}}
\def\nuebar{{\bar{\nu}_e}}
\def\nux{{\nu_x}}
\def\nuxbar{{\bar{\nu}_x}}

\def\nubar{{\bar{\nu}}}
\def\la{\langle}
\def\ra{\rangle}

\def\theta{\vartheta}

\begin{document}

\title[Signatures of supernova shock propagation]
{Neutrino signatures of supernova forward and reverse shock propagation}

\author{R.~Tom\`as, M.~Kachelrie\ss\ and G.~Raffelt}
\address{Max-Planck-Institut f\"ur Physik (Werner-Heisenberg-Institut),
F\"ohringer Ring 6, 80805 M\"unchen, Germany}

\author{A.~Dighe}
\address{Tata Institute of Fundamental Research,
Homi Bhabha Road, Mumbai 400005, India}

\author{H.-T.~Janka and L.~Scheck}
\address{Max-Planck-Institut f\"ur Astrophysik,
Karl-Schwarzschild-Str.~1, 85741 Garching, Germany}

\begin{abstract}
  A few seconds after bounce in a core-collapse supernova, the shock
  wave passes the density region corresponding to resonant neutrino
  oscillations with the ``atmospheric'' neutrino mass difference. The
  transient violation of the adiabaticity condition manifests itself
  in an observable modulation of the neutrino signal from a future
  galactic supernova. In addition to the shock wave propagation
  effects that were previously studied, a reverse shock forms
  when the supersonically expanding neutrino-driven wind 
  collides with the slower earlier supernova ejecta. This
  implies that for some period the neutrinos pass two subsequent
  density discontinuities, giving rise to a ``double dip'' feature in
  the average neutrino energy as a function of time. We study this
  effect both analytically and numerically and find that it allows one
  to trace the positions of the forward and reverse shocks. We show that  
  the energy dependent neutrino conversion probabilities allow one to
  detect oscillations even if the energy spectra of different 
  neutrino flavors are the same as long as the fluxes differ.
  These features are observable in the $\bar\nu_e$ signal
  for an inverted and in the  $\nu_e$ signal for a normal neutrino
  mass hierarchy, provided the 13-mixing angle is ``large''
  ($\sin^2\theta_{13}\gg 10^{-5}$).  
\end{abstract}


\section{Introduction}                                   \label{intro}

While galactic supernovae are rare, the proliferation of existing or
proposed large neutrino detectors has considerably increased the
confidence that a high-statistics supernova (SN) neutrino signal will
eventually be observed. The scientific harvest would be immense. Most
importantly for particle physics, the detailed features of the
neutrino signal may reveal the nature of the neutrino mass ordering
that is extremely difficult to determine experimentally.

On the other hand, a detailed measurement of the neutrino signal from
a galactic SN could yield important clues of the SN explosion
mechanism. Neutrinos undoubtedly play a crucial role for the SN 
dynamics, and neutrino energy deposition behind the SN shock is
able to initiate and power the SN
explosion~\cite{Wi85,BW85}. It is, however, still unclear whether this
energy deposition is indeed sufficiently strong, and current
state-of-the-art models still have problems to produce robust
explosions with
the observed energies (cf., e.g., Ref.~\cite{BRJK03}). Empirical
constraints on the physics deep inside the SN core would
therefore be extremely useful, and neutrinos are the only way for a
direct access besides gravitational waves~\cite{Mueller:2003fs}. 

The neutrinos emitted by the collapsed SN core will pass through the
mantle and envelope of the progenitor star and on the way encounter a
vast range of matter densities $\rho$ from nearly nuclear at the
neutrinosphere to that of interstellar space. The Wolfenstein
effect~\cite{Wolfenstein:1977ue} causes a resonance in neutrino
oscillations~\cite{Mikheev:gs} when 
$\Delta m^2_\nu\cos2\theta/2E_\nu=\pm\sqrt{2}G_{\rm F}Y_{\rm e}\rho$, where
the plus and minus sign refers to neutrinos $\nu$ and antineutrinos
$\bar\nu$, respectively. Therefore, depending on the sign of $\Delta
m^2_\nu$, the resonance occurs in the $\nu$ or the  $\bar\nu$
channel~\cite{Langacker}. For the ``solar'' neutrino mass-squared 
difference of
$\Delta m^2_{21}\approx 81~{\rm meV}^2$ \cite{Kamland} one refers to
the ``L-resonance'' (low density) while for the ``atmospheric'' one of
$|\Delta m^2_{32}|\approx 2300~{\rm meV}^2$ \cite{Maltoni:2004ei} to
the ``H-resonance'' (high density).  The resonance is particularly
important for 
13-oscillations because the 13-mixing angle is known to be small so
that the classic MSW enhancement of flavor conversion in an adiabatic
density gradient is a crucial feature~\cite{Dighe:1999bi}.

\begin{figure}[ht]
\begin{center}
\epsfig{file=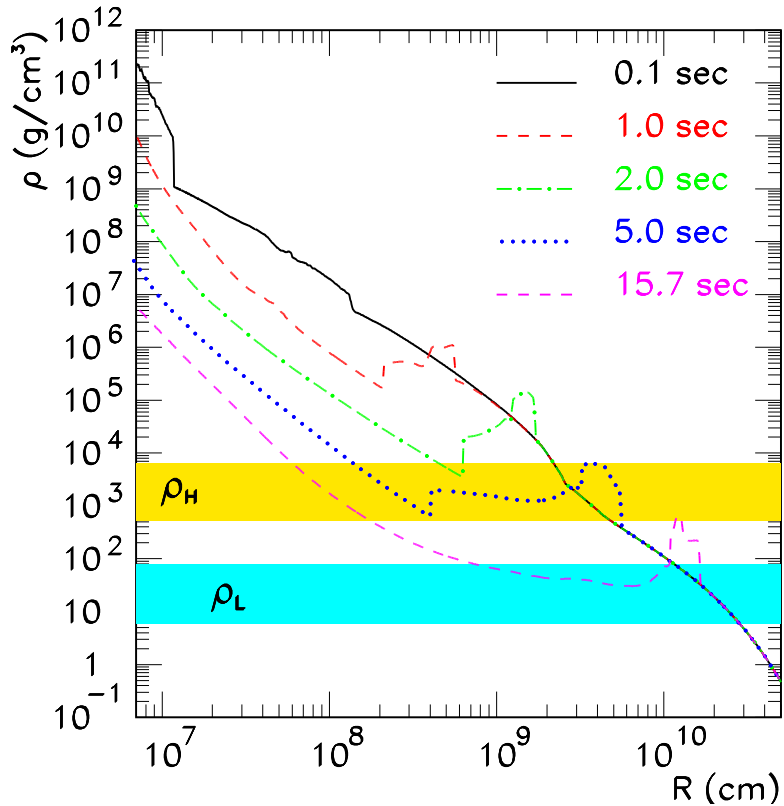,width=4.5in,angle=0}
\end{center}
\label{fig:shockpropagation}
\caption{Shock and reverse-shock propagation in our numerical
calculation. The density profile is shown at the indicated instances
after core bounce. The density region $\rho_{\rm H}$ corresponds to
resonant neutrino oscillations with the atmospheric mass difference,
$\rho_{\rm L}$ to the solar one. The width of the bands reflects the
expected energy range of SN neutrinos.}
\end{figure}

The passage of the shock wave through the density of the H-resonance a
few seconds after core-bounce (Fig.~\ref{fig:shockpropagation}) may 
break adiabaticity, thereby modifying the spectral
features of the observable neutrino flux. Therefore, it is conceivable
that a future large neutrino detector can measure a small modulation
of the neutrino signal caused by the shock-wave propagation, an effect
first discussed by Schirato and Fuller in a seminal
paper~\cite{Schirato} and elaborated by a number of subsequent
authors~\cite{Takahashi:2002yj,Lunardini:2003eh,Fogli:2003dw}. 
Since the density of the
H-resonance depends on energy, the observation of such a modulation
in different neutrino energies would allow one to trace the shock
propagation. On the other hand, the occurrence of this effect
depends on the sign of $\Delta m^2_{31}$ and the value of
$\theta_{13}$, so that observing it in the $\bar\nu_e$ spectra, the
experimentally most accessible channel, would
imply that the neutrino mass ordering is inverted and that
$\sin^2\theta_{13}\gg 10^{-5}$.

We here explore a new feature of the shock-wave ``fingerprint'' in
the neutrino signal. Some time after the onset of the explosion
a neutrino-driven baryonic wind develops and collides with the
earlier, more slowly expanding supernova ejecta. A reverse shock
is seen to form in all models which were computed with sufficient
resolution and seems to be a generic feature, although
the exact propagation history depends on the detailed
dynamics during the early stages of the supernova explosion.
Moreover, violent convective instabilities and large anisotropies 
are observed in the neutrino-heated layer behind the SN shock
in multi-dimensional SN simulations and
imply a significant angular variation of the
instantaneous density profiles even in a single star. Still, the
simultaneous propagation of a direct and an inverse shock wave imply
that for some time the neutrinos may encounter two subsequent density
discontinuities, leading to significantly different spectral features
than are caused by a single crossing.

Numerical SN simulations do not yet lead to robust explosions so that
it remains unclear if an important physical ingredient is missing in
our understanding of the SN phenomenon. Even assuming that only 
minor aspects need to be tweaked, this situation implies that 
hydrodynamic state-of-the-art models of successful explosions 
with self-consistently determined neutrino fluxes
and spectra do not exist at present. Therefore, we limit our
discussion to a simple analytic study with a schematic density profile
and an example of a detailed numerical model, using for both cases 
schematic primary neutrino fluxes and spectra that are representative
for neutrino spectra discussed in the literature. In this
way we identify a generic ``double dip'' feature caused by the
presence of two density discontinuities.
We find that the position of the two dips in time can be connected to the
positions of the forward and reverse shock. The observation of the
dips in different energy bins allows therefore not only the
identification of the reverse shock but also to trace the
propagation of both shocks.
Since the neutrino conversion probabilities are energy dependent 
during the passage of the shocks through the H-resonance, neutrino
oscillations can be detected even if the energy spectra of different
neutrino flavors have the same shape but different luminosities.

We begin in Sec.~2 with an explanation of the reverse shock formation
and a discussion of our numerical shock-propagation examples. In
Sec.~3 we use a schematic model of the density profile to derive
analytically the generic features imprinted on the observable neutrino
signal. In Sec.~4 we use a concrete numerical example of shock and
reverse shock propagation to calculate their typical signatures in a large
$\bar\nu_e$ detector. In Sec.~5 we discuss and summarize our findings.

\section{Our numerical SN model}
\label{reverse}

Relying on the viability of the neutrino-heating mechanism we
have performed simulations of supernova explosions that followed 
the evolution in spherical symmetry until more than 25 seconds 
after shock formation. In order to include the effects of convection
we have also done runs in two dimensions for about one second of
post bounce evolution. Because of the persistent problems of 
complete models in obtaining successful explosions (see, e.g., 
Ref.~\cite{BRJK03}), we replaced the high-density interior of
the contracting nascent neutron star by a time-dependent inner 
boundary where the neutrino luminosities and spectra were 
imposed such that the neutrino energy transfer to the shock was 
sufficiently strong for shock revival (for more details,
see Refs.~\cite{SPJKM04,Sc04}).

The density profiles used for the evolving stellar background 
in the present work (cf.\ Fig.~\ref{fig:shockpropagation})
were taken from simulations of a 15~$M_{\odot}$ 
progenitor~\cite{WW95}.
Although the employed neutrino parameters are in the ballpark
of results from elaborate neutrino transport calculations 
in supernovae, detailed information about neutrino fluxes and
spectra that are consistent with explosion models and with the
cooling history of the newly formed neutron star are currently
not available. For this reason we consider the density profiles
as exemplary for realistic supernova conditions and combine them
with a schematic description of the neutrino emission that is
supposed to contain generic features of the expected neutrino 
signal.

\begin{figure}
\begin{center}
\epsfig{file=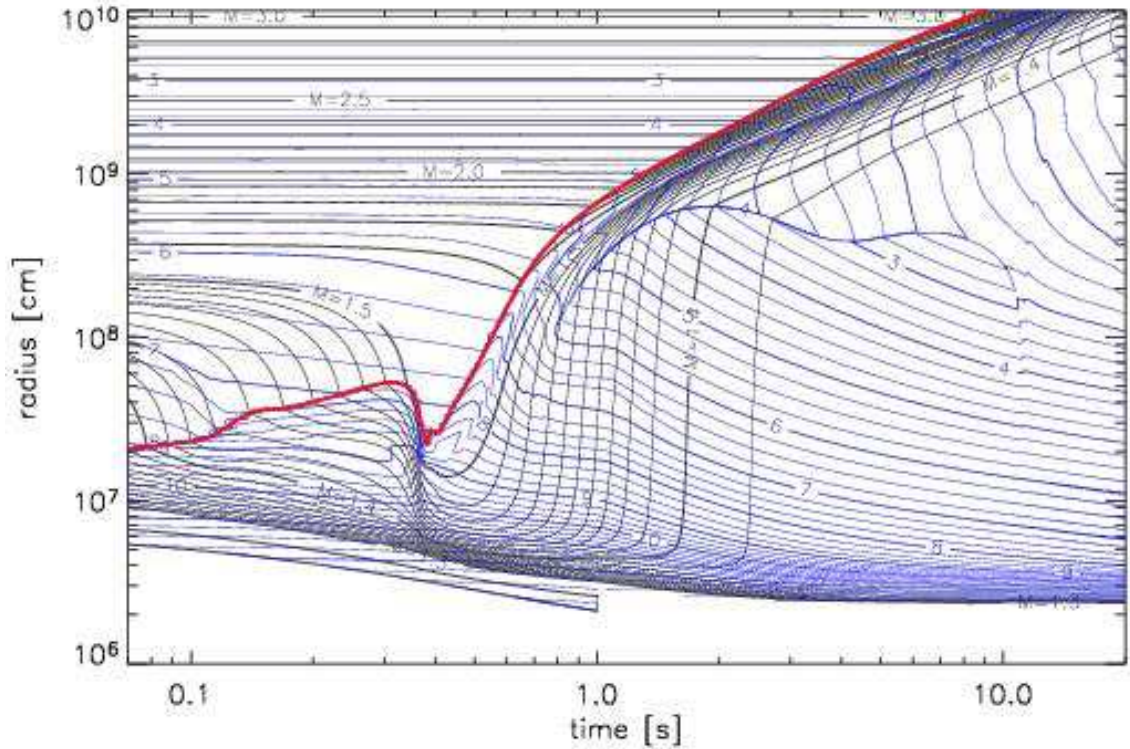,width=10.0truecm,angle=90}
\caption{Early evolution (for times between 0.07 and 20 seconds 
after shock formation) of the supernova explosion 
(with an energy of about $1.5\times 10^{51}$~erg) of a 
15~$M_{\odot}$ progenitor according to a spherically symmetric 
simulation in which the contracting neutron star was replaced by 
an inner boundary that acts as a neutrino source 
sufficiently intense to cause a neutrino-driven explosion.
The black lines mark the space-time
trajectories of selected mass shells, the blue lines represent
corresponding contours of constant density, and the thick red
line indicates the position of the supernova shock. The reverse
shock and the contact discontinuity can be recognized from kinks
and a compression of the isodensity curves in the region behind 
the forward shock.}
\end{center}
\label{fig:trajectories}
\end{figure}

\begin{figure}
\begin{center}
\epsfclipon \epsfig{file=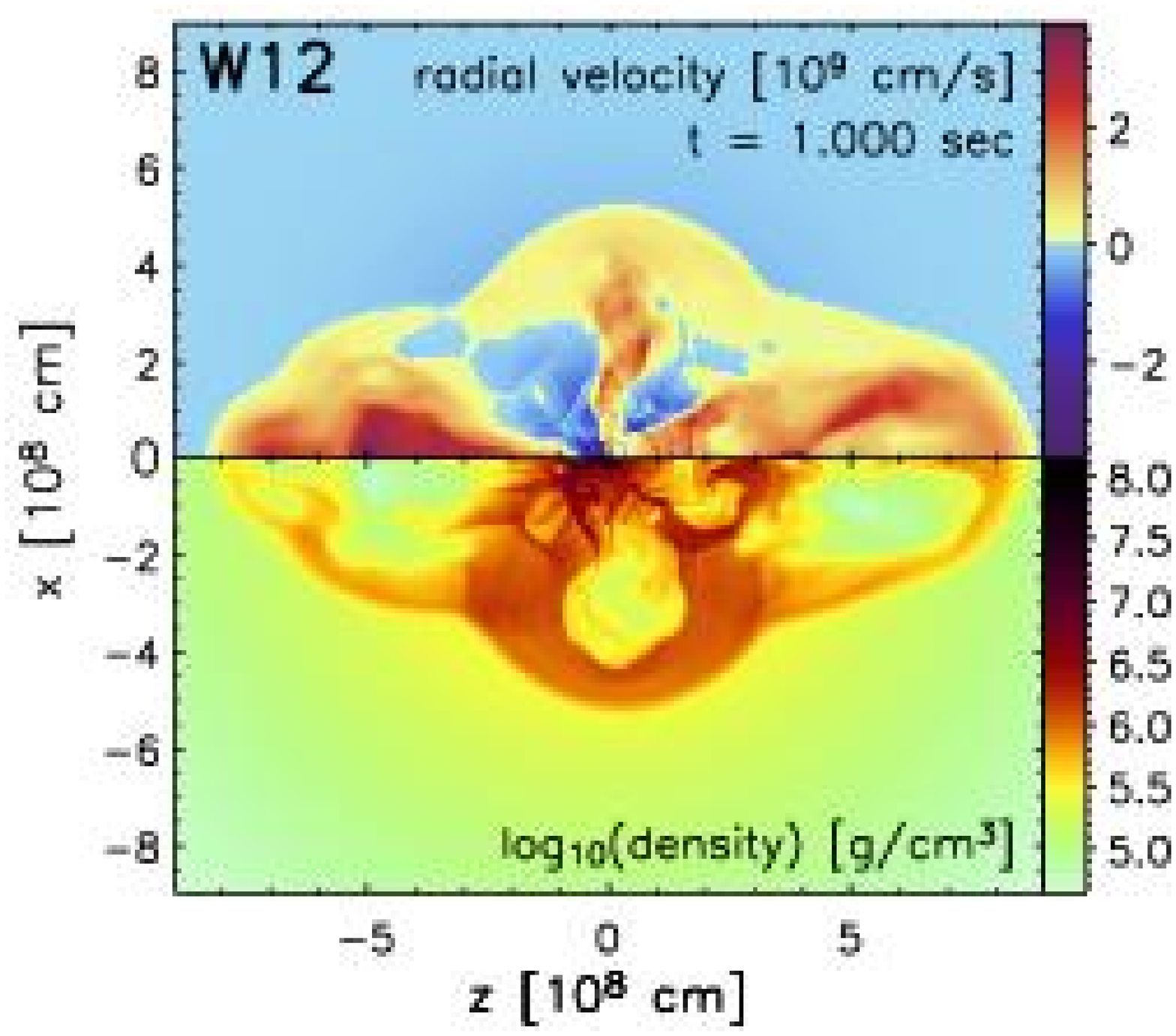,height=6.0truecm,angle=0}
\epsfclipon \epsfig{file=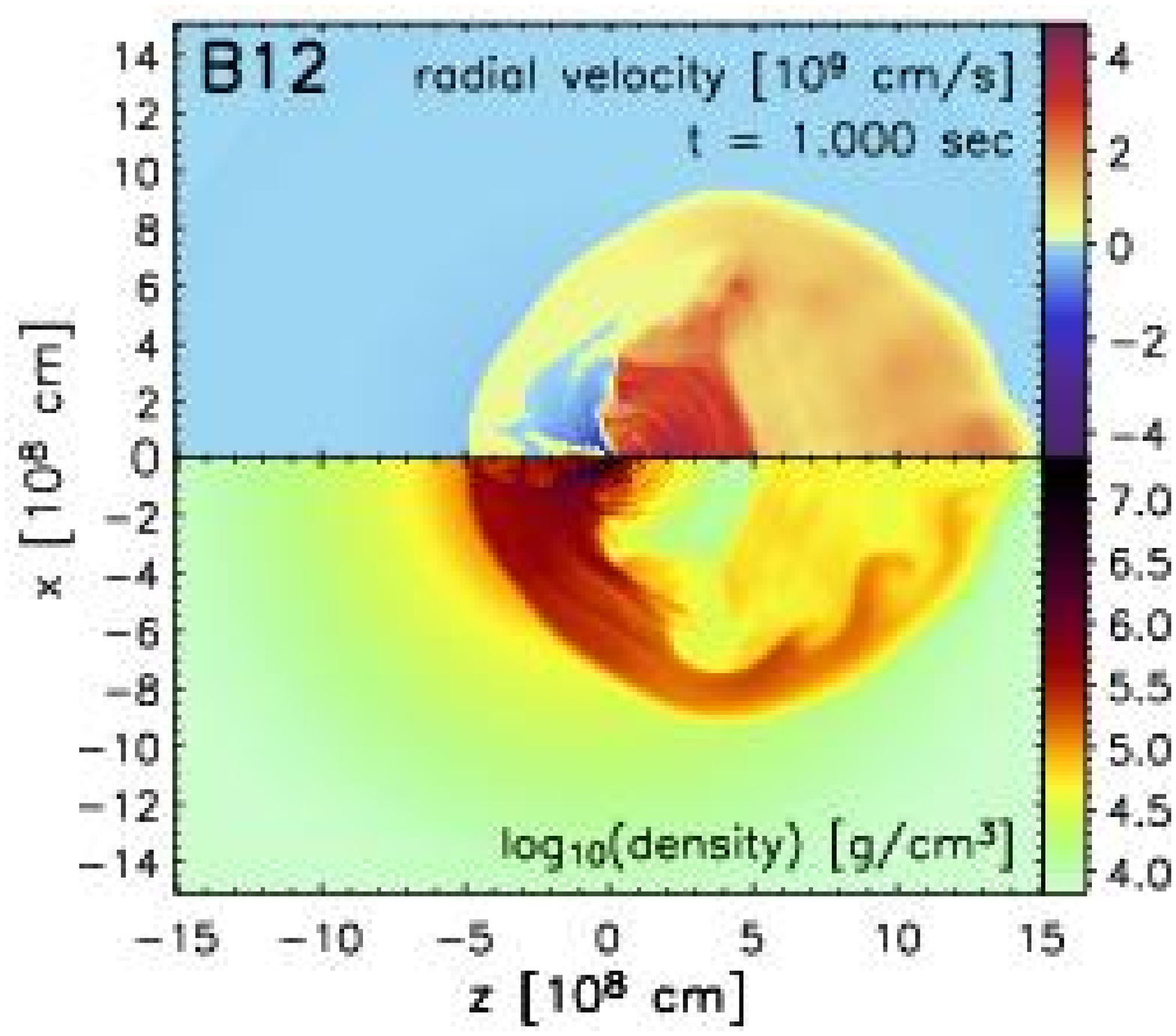,height=6.0truecm,angle=0}\\
\epsfclipon \epsfig{file=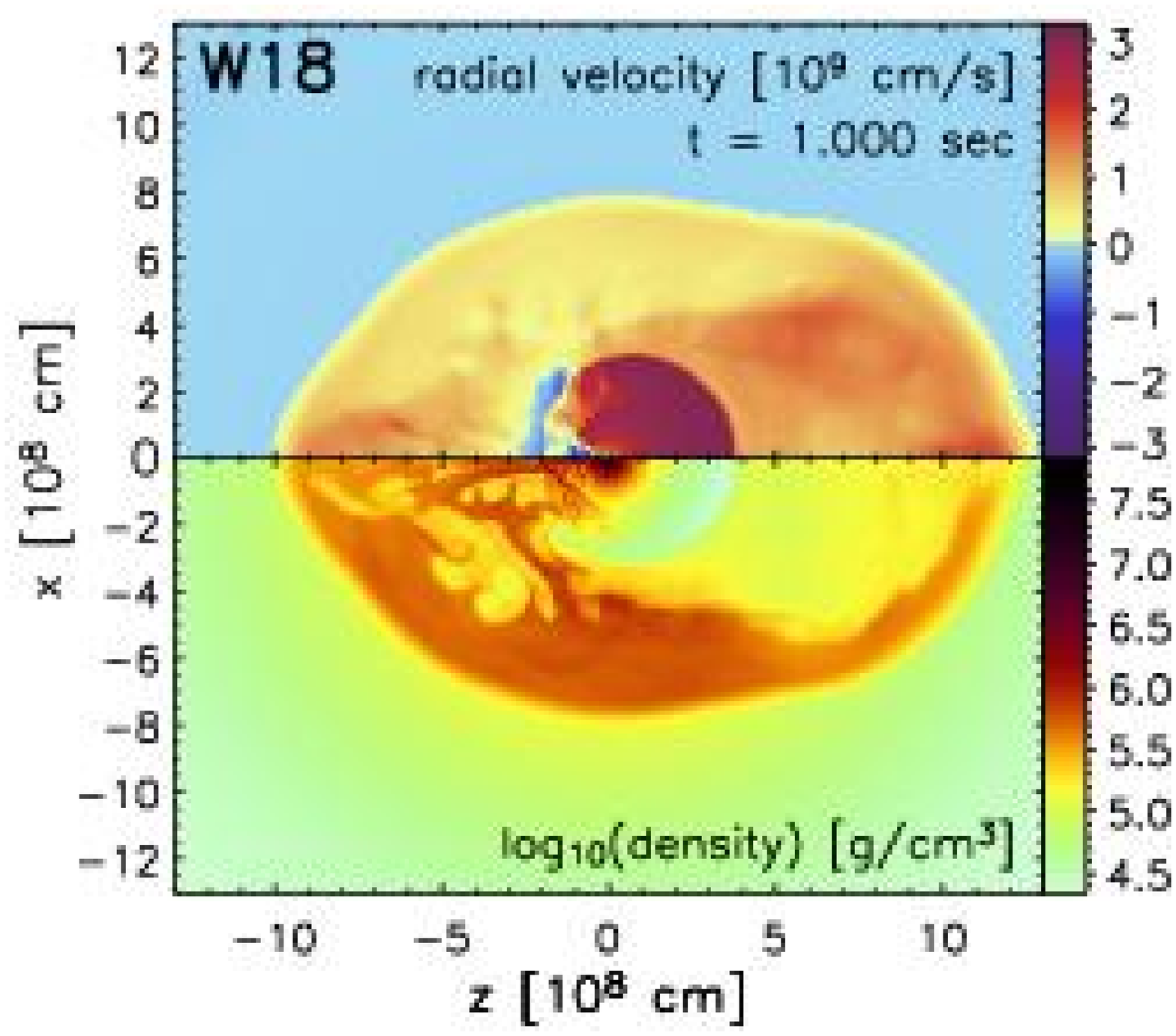,height=6.0truecm,angle=0}
\epsfclipon \epsfig{file=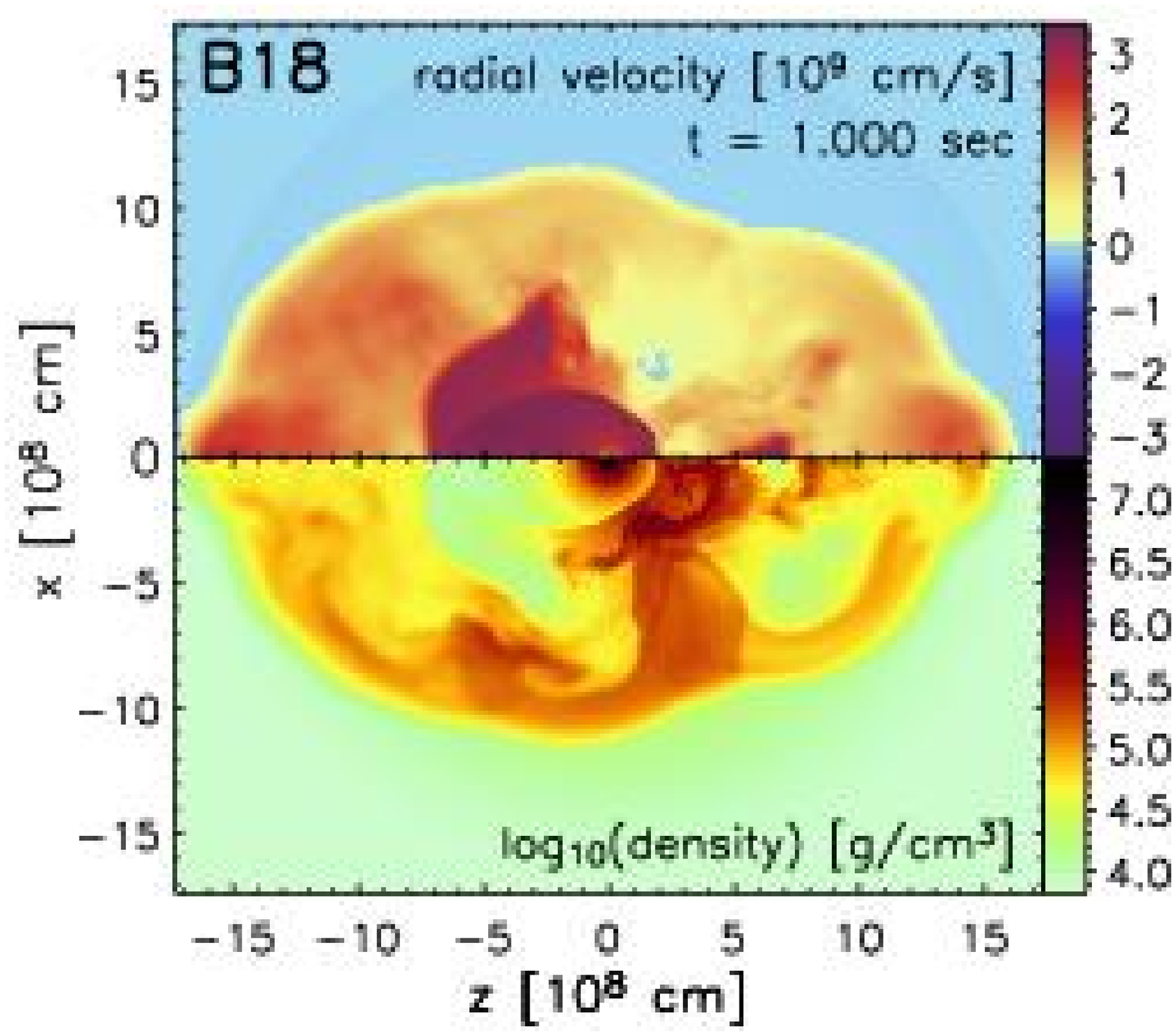,height=6.0truecm,angle=0}
\caption{Radial velocity (upper half panels) and 
logarithm of density (lower half panels)
color coded for four different two-dimensional simulations at one
second after supernova shock formation. The explosion energies of
these models are about $0.3\times 10^{51}$~erg, 
$0.4\times 10^{51}$~erg, $0.9\times 10^{51}$~erg, and
$1.2\times 10^{51}$~erg (from top left to bottom right),
respectively. The postshock structure is much more complex than
in spherical symmetry, and depends strongly on the direction. 
While the contact discontinuity is distorted by hydrodynamic
instabilities and marks the boundary between low-density
neutrino-heated bubbles and matter that has been accreted
and accelerated by the supernova shock, the reverse shock is a 
clear signature of the onset of a supersonically expanding 
neutrino-driven wind which 
runs into the earlier, slower supernova ejecta. But also
multiple (reverse) shocks can occur due to collisions between
downflows and neutrino-heated, expanding plasma.}
\end{center}
\label{fig:twodimmodels}
\end{figure}

The space-time evolution of the explosion of this 15~$M_{\odot}$
star in a spherically symmetric simulation  (i.e., neglecting the
effects of hydrodynamic instabilities) can be seen in 
Fig.~\ref{fig:trajectories}. The shock expands from the point of
its formation (not shown in this figure that displays only times
later than 70~ms after core bounce) to a radius of about 
100~km where it converts into a standing accretion shock. 
Its subsequent slow outward motion reflects the quasi-stationary 
adjustment of its position in response to the decreasing rate of
mass infall from the outer stellar layers and to the 
energy deposition by
neutrinos in the ``gain region'' behind the shock. At a radius
of about 500~km the shock reaches a temporary maximum and
starts retreating again because of decaying neutrino luminosities 
and a corresponding drop of the neutrino-heating rate. The 
subsequent sharp increase of the neutrino luminosities and of the
energy deposition behind the shock 
is sufficient to trigger a successful explosion ``in a 
second attempt.'' This behavior is typical of one-dimensional
simulations with conditions close to the threshold for
shock revival.

On its way out the shock reverses the infall of the 
swept-up matter. Continuous neutrino energy transfer
starts an outward acceleration of heated material in the
gain layer around the neutron star. At the interface between
this ``neutrino-driven wind'' to the outer ejecta a
``contact discontinuity'' is formed. It is visible as
a strip of compressed density contours somewhat behind the
outgoing supernova shock in Fig.~\ref{fig:trajectories}. 
Even farther behind the forward shock, the neutrino-driven
wind, whose velocity increases rapidly with distance from the 
neutron star, collides with more slowly moving material and
is decelerated again. The strongly negative velocity gradient
at this location steepens into a reverse shock when the wind 
velocity begins to exceed the local sound speed~\cite{JM95}. 
In the model of Fig.~\ref{fig:trajectories} this happens at 
$t\gsim\,$800$\,$ms post bounce at a radius 
$r\gsim\,$1000$\,$km.

The forward and reverse shocks are sharp discontinuities
where density, pressure (as all other state variables) 
and velocity of the stellar plasma
change on the microscopic (sub-millimeter) scale of the ion 
mean free path, on which the dissipation of kinetic to thermal
energy in the shock occurs. In contrast, the ``contact 
discontinuity'' is characterized by a density jump
but constant velocity and pressure. It
emerges in the matter in the transition 
region between shock-accelerated and neutrino-heated supernova
ejecta. Its width is therefore a result of the explosion dynamics
and found to be 200--300$\,$km in our one-dimensional model.
However, because of our exponentially coarsening grid we are 
able to resolve this outward moving feature only for 2--3 
seconds and are unable to make predictions of its exact 
structure at later times. A schematic representation of
the described structure is given in Fig.~\ref{cartoon}.

The multi-dimensional situation is much more complex 
(cf. Fig.~\ref{fig:twodimmodels}). The forward shock can now
be highly deformed, and the reverse shock,
which still separates the supersonically expanding 
neutrino-driven wind from subsonically moving outer ejecta,
shows large angular variation. This reflects the anisotropy
of the fluid flow between the shock and the neutron star, which 
is characterized by narrow downflows of cool gas and rising 
bubbles of low-density, neutrino-heated matter. These 
convective mass motions 
create a highly distorted structure in which multiple shocks
can occur due to collisions of downflows with expanding, hot
matter. Contact discontinuities
follow the interface between the rising hot bubbles
and shock accelerated plasma that has lower entropy and higher 
density. Their shape is affected by Rayleigh-Taylor and 
Kelvin-Helmholtz instabilities.
Since the explosion is highly asymmetric, the 
exact structure varies strongly with the direction. Although the
generic features of the one-dimensional situation are retained,
the details of the multi-dimensional structure evolve chaotically
from random initial perturbations and cannot be uniquely
determined by models.

\section{Effect of the shock waves on neutrino propagation}
\label{analytical}

In present and planned water Cherenkov and scintillation detectors 
the main  neutrino detection channel is the inverse beta decay
reaction, $\bar\nu_e p\to ne^+$, that allows also for the reconstruction
of the neutrino energies. Therefore we consider only the $\bar\nu_e$
spectrum in the following.  
However an analogous analysis can be easily performed in the neutrino
channel for a detector able to measure the $\nue$ spectrum, using
e.g. liquid argon.   

The antineutrino spectra arriving at the Earth are 
determined by the primary antineutrino spectra as well as
the neutrino mixing scenario,
\be
F_\nuebar(E) = \bar{p}(E) F_\nuebar^0(E) + [1-\bar{p}(E)] F_\nuxbar^0(E) \,,
\label{pbar-def}
\ee
where $\bar{p}(E)$ is the survival probability of a $\bar\nu_e$ with
energy $E$ after propagation through the SN mantle%
\footnote{We neglect generally Earth matter effects. They slightly
  increase the chances to detect the SN shock propagation, cf. figure~8.},  
the superscript zero denotes the primary fluxes, and 
$\nuxbar$ stands for either $\nubar_\mu$ or $\nubar_\tau$.
In general, all quantities of Eq.~(\ref{pbar-def}) are not only energy
but also time dependent. In particular the survival probability $\bar{p}(E)$ 
becomes time dependent during the passage of the shocks through the
H-resonance. Before we discuss the effect of the shock propagation on
the neutrino spectra, we first recall the case of three-flavor neutrino
oscillations in a static density profile. In this case, $\bar{p}(E)$
is not only constant in time, but also independent of energy if
$\sin^2\theta_{13}\lsim 10^{-5}$ or $\sin^2\theta_{13}\gsim 10^{-3}$. 
Since about 60\% of the observed neutrinos will be detected
in the first two seconds after bounce, i.e. before the shock reaches
the H-resonance of low-energy neutrinos, it is for many purposes
sufficient to perform the simpler analysis of neutrino propagation
through the static envelope of the progenitor. For us, the static case
discussed in the next subsection serves as a reference to measure the
modulations induced by the propagating shock.

\subsection{Static case before the arrival of the shock wave at the
  H-resonance}

The survival probability is determined by the flavor conversions
that take place in the resonance layers, where 
$\rho_{\rm res} \approx m_{\rm N} \dmsq_{i}\cos 2\theta / 
(2 \sqrt{2} G_{\rm F} Y_{\rm e} E)$. Here $\dmsq_{i}$ and $\theta$ are the
relevant mass difference and mixing angle of the neutrinos,
$m_{\rm N}$ is the nucleon mass,
$G_{\rm F}$ the Fermi constant and $Y_{\rm e}$ the electron fraction.
In contrast to the solar case, SN neutrinos must pass through
two resonance layers: the H-resonance layer at 
$\rho_{\rm H}\sim 10^3$~g/cm$^3$ corresponds to $\Delta m^2_{\rm atm}$,
whereas the L-resonance layer at 
$\rho_{\rm L}\sim 10$~g/cm$^3$ corresponds to $\Delta m^2_{\odot}$.
This hierarchy of the resonance densities, along with their
relatively small widths, allows the transitions in the two 
resonance layers to be considered independently~\cite{Dighe:1999bi}.

The neutrino survival probabilities can be characterized by the degree
of adiabaticity of the resonances traversed, which are directly 
connected to the neutrino mixing scheme. 
In particular, whereas the L-resonance is always adiabatic and 
appears only in the neutrino channel, the adiabaticity of
the H-resonance depends on the value of $\theta_{13}$, and 
the resonance shows up in the neutrino or antineutrino channel 
for a normal or inverted mass hierarchy respectively.
Table~\ref{tab-pbar} shows the survival probabilities for neutrinos,
$p$, and antineutrinos, $\bar{p}$, in various mixing scenarios. 
For intermediate values of $\theta_{13}$, i.e.
$10^{-5}\lsim\sin^2 \theta_{13} \lsim 10^{-3}$,
the survival probabilities depend even without shock on energy as well
as the details of the density profile of the SN. 

For large values of $\theta_{13}$ and the static density profile of
the progenitor, the H-resonance is adiabatic. In the case of normal
mass hierarchy, scenario A, the resonance takes place in the neutrino
channel, and antineutrinos are not affected. 
For an inverted mass hierarchy, scenario B, the resonance occurs 
in the antineutrino channel, so that practically all the primary 
$\nuebar$ are converted to $\bar{\nu}_3$ and arrive at the Earth as
$\nuxbar$. This corresponds to an almost complete interchange of
$\nuebar$ and $\nuxbar$ spectra. In scenario C, the resonance is
strongly non-adiabatic, and hence ineffective. 
In both the scenarios A and C, the primary $\nuebar$ leave the star
as $\bar\nu_1$, which implies a partial mixing between $\nuebar$ and
$\nuxbar$ spectra.

\begin{table}
\begin{center}
\begin{tabular}{llccc}
\hline
Scenario & 
Hierarchy &  $\sin^2 \theta_{13}$  &  $p$ &  $\bar{p}$ \\
\hline
A & Normal & $\gsim 10^{-3}$  & 0  & $\cos^2\theta_\odot$ \\
B & Inverted &  $\gsim 10^{-3}$ &  $\sin^2\theta_\odot$ &  0 \\
C & Any & $\lsim 10^{-5}$  & $\sin^2\theta_\odot$ 
&  $\cos^2\theta_\odot$ \\
\hline
\end{tabular}
\caption{
Survival probabilities for neutrinos, $p$, and
antineutrinos, $\bar{p}$, in various mixing scenarios for a static
density profile.
\label{tab-pbar}}
\end{center} 
\end{table}

\subsection{Shock waves passing through the H-resonance} 

In our numerical model, the outgoing shock wave reaches the
H-resonance layer around two seconds after bounce and the L-resonance
layer around ten seconds after bounce. Since for $t\gsim 11$~s the
H-resonance is adiabatic, neutrinos created as $\bar\nu_e$ arrive in
case B as $\bar\nu_3$ at the L-resonance. Thus, the L-resonance that
mixes 12-states does not affect the survival probability $\bar p$,
independent of the passage of shock waves. In contrast, it depends on
the neutrino mixing scenario,  whether the shock waves modulate the
neutrino spectra during their passage through the H-resonance: 
The cases A and C will not show any evidence of shock wave propagation
in the observed $\nuebar$ spectrum, either because there is no
resonance in the antineutrino channel as in scenario A, or because the
resonance is always strongly non-adiabatic as in scenario C.
However, in scenario B, the sudden change in the density breaks  the
adiabaticity of the resonance, leading to observable consequences in the
$\nuebar$ spectrum. Let us discuss this scenario henceforth.

\begin{figure}
\begin{center}
\epsfig{file=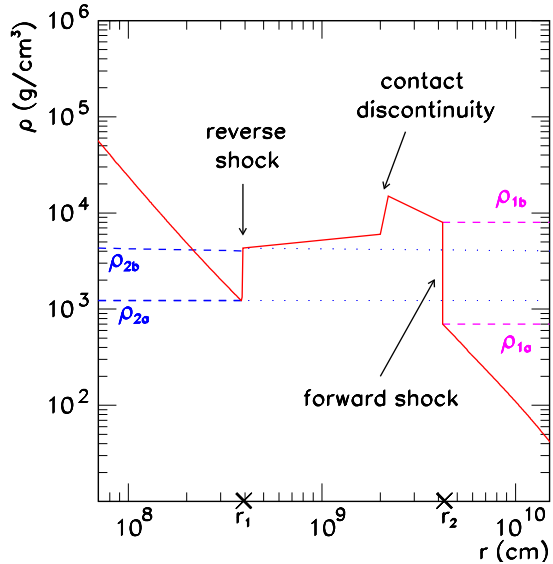,width=8cm}
\end{center}
\caption{Schematic density profile in the presence of a forward and
  reverse shock wave. Between forward and reverse shock a contact
  discontinuity can appear. 
\label{cartoon}}
\end{figure}

In order to analytically understand the signatures of the shock wave 
in the observed $\nuebar$ spectrum, let us first assume that
the shock wave causes the adiabatic H-resonance to become 
completely non-adiabatic at the forward shock front and the reverse
shock front, while at all other places the resonance remains
adiabatic. In the snapshot of the shock wave shown in
Fig.~\ref{cartoon}, non-adiabatic transitions take place 
only at the two radii $r_1$ and $r_2$, where the forward and
the reverse shock waves are present at that time.

\begin{figure}
\begin{center}
\epsfig{file=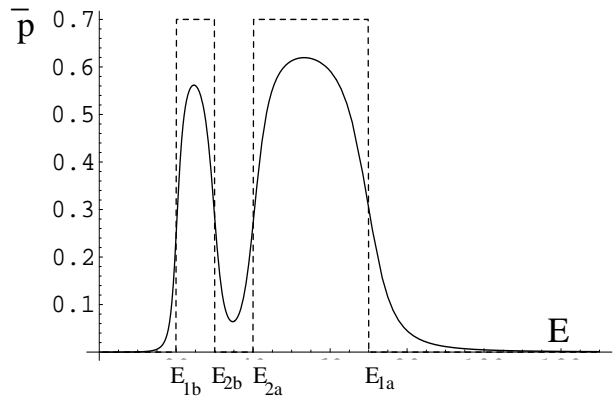,width=8cm}
\end{center}
\caption{Survival probability $\bar{p}$ as a function of energy $E$
  assuming $|\theta_a-\theta_b|\approx\pi/2$ (dotted line) or using
  Eq.~(\ref{exact-pbar}) (solid line) with $\cos^2 \theta_\odot=0.7$. 
\label{squarewave}}
\end{figure}

Neutrinos with energy $E$ undergo H-resonance in the region of 
density 
\be
 \rho = \frac{m_{\rm N}\dmsq_{\rm atm}\cos2\theta_{13}}{2 \sqrt{2} G_{\rm F} Y_{\rm e} E} 
 \approx 
 600\,{\rm g/cm}^3 \; \cos2\theta_{13}\:\frac{25~{\rm MeV}}{E}\:\frac{1}{Y_{\rm e}}
\,.
\label{rescondition}
\ee
The maximum and minimum densities at the two shock fronts,
as illustrated in Fig.~\ref{cartoon}, then directly give 
the range of energies that undergo a non-adiabatic resonance at 
these shock fronts. We introduce now the two approximations of our
toy-model used in this subsection: First, we assume that all
resonances are either completely adiabatic or completely
non-adiabatic. Second, we require that the density jump at the
resonances is large, i.e.\ that the change of the medium mixing angle
is close to $\pi/2$. Then the survival probability $\bar{p}(E)$ is a
rectangular wave as a function of energy, as shown in
Fig.~\ref{squarewave} with a dotted line.
Clearly, as the shock wave propagates to lower densities, the
``turning points'' of the square wave shift upwards in energy.
The amplitude of the square wave, $\cos^2 \theta_\odot$,
is fixed by the value of the solar mixing angle.
We can relax the second approximation and use the result that,
if the density changes sharply from $\rho_b$ to $\rho_a$,
the survival probability for $\nuebar$ is 
\be
\bar{p} = \sin^2(\theta_{a}-\theta_{b}) 
\cos^2 \theta_\odot \,,
\label{exact-pbar}
\ee
where $\theta_{a}$ and $\theta_{b}$ are the values of
$\theta_{13}$ in matter at the densities $\rho_a$ and $\rho_b$, 
respectively. Using this, the rectangular wave for $\bar{p}$
gets modified as shown in Fig.~\ref{squarewave} with a solid line.
However, in order to get a simple understanding of the new features
of the conversion probability  we will use the rectangular wave for
$\bar p$ in the rest of this section.

The primary spectra of neutrinos can be parameterized
by~\cite{MKeil,Keil:2002in} 
\be
 F^0_i(E) =
 \frac{\Phi_i}{\langle E_i \rangle}\,
 \frac{\beta_i^{\beta_i}}{\Gamma(\beta_i)}  
 \left(\frac{E}{\langle E_i \rangle}\right)^{\beta_i-1} 
 \exp\left(-\beta_i\frac{E}{\langle E_i \rangle}\right) \,,
\label{flux-form}
\ee
where $\langle E_i \rangle$ denotes their average energy,
$\beta_i$ is a dimensionless parameter that relates to the
width of the spectrum and typically takes on values 3.5--6, depending
on the flavor and the phase of neutrino emission.  
Let us approximate the energy dependence of the
cross section of the inverse beta decay reaction by $E^2$, and assume
an ideal detector. Using Eq.~(\ref{pbar-def}),
Table~\ref{tab-pbar} and the survival probability as shown in
Fig.~\ref{squarewave}, the number of $\nuebar$ events observed
in a certain time interval is
\be
 N_{\rm obs} = {\cal N} \left[
\Phi_\nuxbar \frac{ \langle E_\nuxbar \rangle^2}{\beta_\nuxbar^2}
\frac{\Gamma(\beta_\nuxbar+2)}{\Gamma(\beta_\nuxbar)}  +
\cos^2 \theta_\odot (\Phi_\nuebar g_2^\nuebar -
\Phi_\nuxbar g_2^\nuxbar ) \right] \, ,
\label{n-def}
\ee
where ${\cal N}$ is a normalization constant, $\Gamma(z)$ is
the Euler gamma function, and we have introduced the functions
\be 
g_k^i = \frac{\langle E_\nuxbar \rangle^k}{\beta_i^k \Gamma(\beta_i)}
\left[  \Gamma \left( a, \frac{E_{1}}{\langle E_i \rangle} \beta_i, 
\frac{E_{2}}{\langle E_i \rangle} \beta_i \right) + 
 \Gamma \left( a, \frac{E_{3}}{\langle E_i \rangle} \beta_i, 
\frac{E_{4}}{\langle E_i \rangle} \beta_i \right)
\right] \,.
\label{g-def}
\ee
Here, $\Gamma(a,b, c)=\int_b^c dt\: t^{a-1}e^{-t}$ is a generalized
incomplete Gamma function, and $a=\beta+k$. 
The values $E_1, E_2, E_3, E_4$ are the same as 
$E_{1a}, E_{1b}, E_{2a}, E_{2b}$ in ascending order.
The $m$th moment of the total energy $E_{\rm obs}^m$ deposited in the 
detector during a certain time interval is similarly given by 
\be
\hskip-1.cm
 E_{\rm obs}^m = {\cal N} \left[
 \Phi_\nuxbar \frac{ \langle E_\nuxbar \rangle^{2+m}}{\beta_\nuxbar^{2+m}}
 \frac{\Gamma(\beta_\nuxbar+2+m)}{\Gamma(\beta_\nuxbar)}  +
 \cos^2 \theta_\odot (\Phi_\nuebar g_{2+m}^\nuebar -
 \Phi_\nuxbar g_{2+m}^\nuxbar ) \right] \,.
\label{lum-def}
\ee

Since the values of $E_{1a}$,  $E_{1b}$, $E_{2a}$, $E_{2b}$ change as
a function of time, so do the values of $N_{\rm obs}, E_{\rm obs}$. 
As time goes on, the shock wave reaches layers with smaller densities
and therefore, according to Eq.~(\ref{rescondition}), higher-energetic
neutrinos are affected by the wave propagation. The time dependence of
the numerical profiles can be imitated starting from $t=1\,$s by 
$E_i(t) = (t/{\rm s})^{a_i}\ln(t/{\rm s})$~MeV, and where 
$a_i=\{1.3,1.5,1.7,2.0\}$. As a consequence,  the double square
structure in the survival probability shown in Fig.~\ref{squarewave}  
keeps on shifting towards higher energies with time.

In order to study the model dependence of our results, we consider
three models that give very different predictions for the neutrino
spectra.  Two of them, G1 and G2, are motivated by the recent Garching
calculation~\cite{garching} that includes all relevant neutrino
interaction effects like nucleon bremsstrahlung, neutrino pair
processes, weak magnetism, nucleon recoils and nuclear correlation
effects.  The third one is from the Livermore
simulation~\cite{livermore} that represents more traditional
predictions for flavor-dependent SN neutrino spectra used in many
previous analyses.  The parameters of these models are shown in
Table~\ref{tab:models}.

\begin{table}[ht]
\begin{indented}\item[]
\begin{tabular}{cccccc}
\hline
Model & $\langle E_0(\nu_e) \rangle$ & $\langle E_0(\nuebar) \rangle$&
$\langle E_0(\nux) \rangle$ & {\large $\frac{\Phi_0(\nu_e)}{\Phi_0(\nu_x)}$} &
{\large $\frac{\Phi_0(\nuebar)}{\Phi_0(\nu_x)}$}\\
\hline
L & 12 & 15 & 24 & 2.0 & 1.6 \\
G1 & 12 & 15 & 18 & 0.8 & 0.8 \\
G2 & 12 & 15 & 15 & 0.5 & 0.5 \\
\hline
\end{tabular}
\end{indented}
\caption{The parameters of the used primary neutrino spectra models
  motivated from SN simulations of the Garching (G1, G2) and the
  Livermore (L) group. We assume $\beta_\nuxbar=4$ and $\beta_\nuebar=5$. 
\label{tab:models}}
\end{table}

\begin{figure}[h]
\epsfig{file=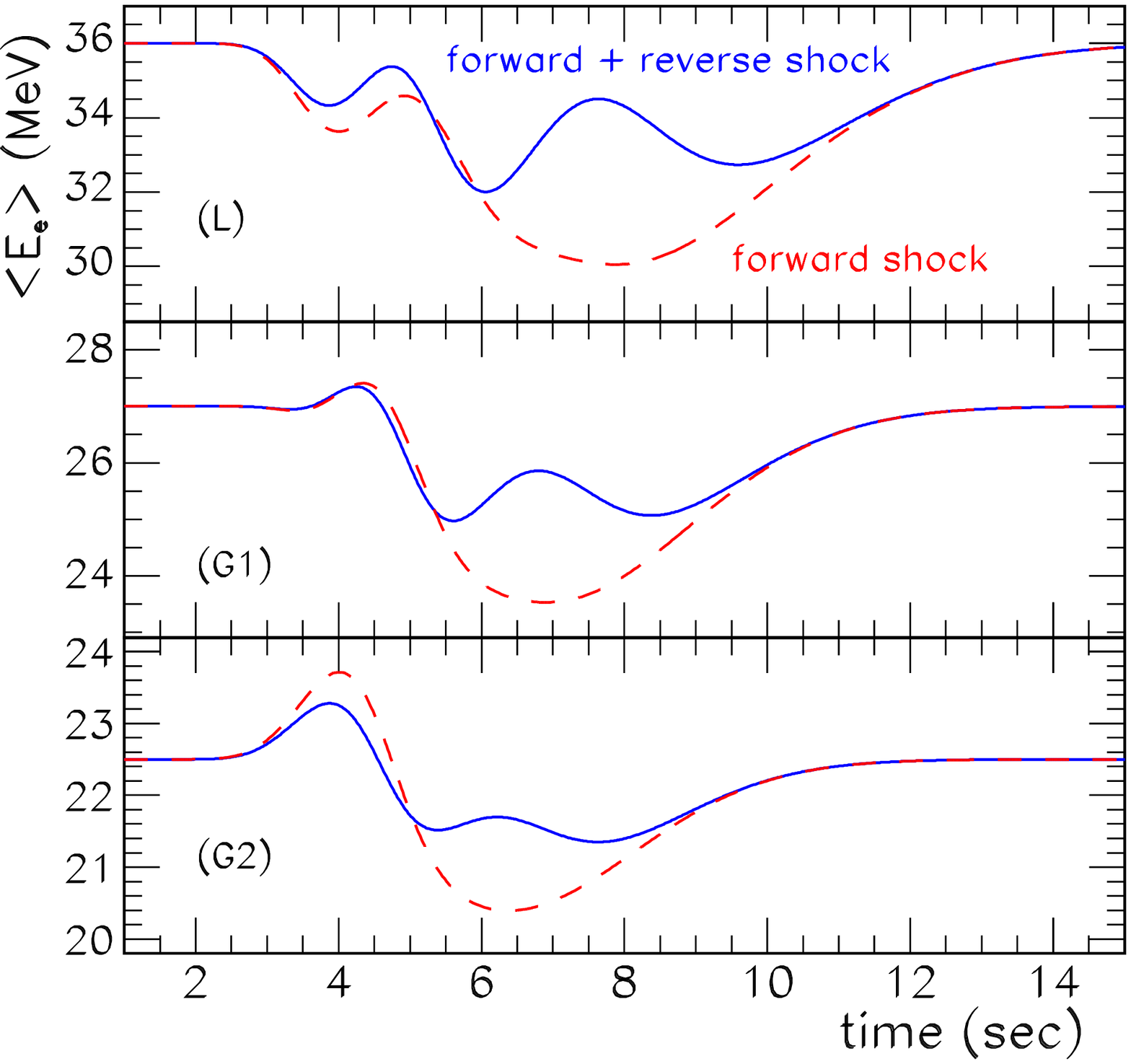,width=3in}
\epsfig{file=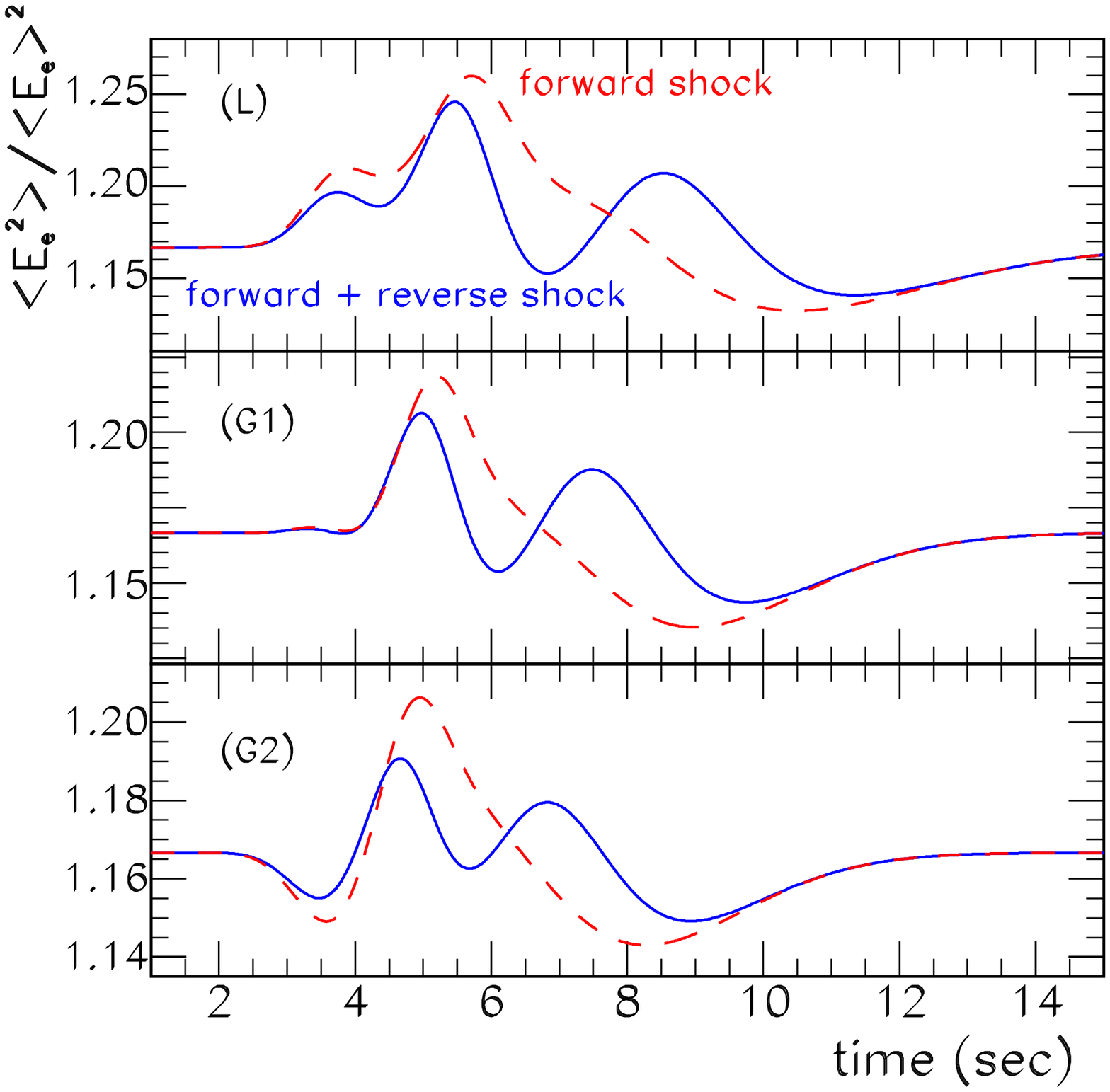,width=3in}
\caption{Time dependences of the observables $\la E_e \ra$
  (left) and  $\la
  E_e^2 \ra / \la E_e \ra^2$ (right) for different primary spectra,
  (L, G1, G2), and
  assuming both forward and reverse shock (blue solid line) and only
  forward shock wave (red dashed line).
\label{time-an}}
\end{figure}

In Fig.~\ref{time-an}, we show the values of the average event energy 
$\la E_e \ra \equiv  E_{\rm obs}/N_{\rm obs}$ observed (left) as well
as $\xi\equiv \la E_e^2 \ra / \la E_e \ra^2$  (right) as functions of time
for the three models of Tab~\ref{tab:models}. The parameter $\xi$
measures the pinching of the observed spectrum and indicates
therefore if it is a superposition of primary spectra with different
shapes. We compare the case where both forward and reverse shock waves
are present with the forward shock only case. We approximate the latter
by setting $E_{2a} = E_{2b}$.  

In the absence of shock waves, the adiabaticity of the propagation in
case B ensures that the $\nuebar$ flux arriving at the Earth,
$F_\nuebar(E)$, basically coincides with $F^0_\nuxbar(E)$. 
Thus, breaking that adiabaticity will imply a partial presence of
$F^0_\nuebar(E)$ in $F_\nuebar(E)$. Therefore any departure of the
observables from their values in the no-shock case induced by the
shock depends strongly on which part of the energy spectra is affected
by the shock wave, and how different the original spectra are at
those energies. 

At early times, $t\lsim 4$~s, the shock affects only the low-energy part of
the spectra, $E \lsim 20$~MeV. At these energies, the ratio
$F^0_\nuxbar(E)/F^0_\nuebar(E)$ varies for different SN models: it is
smaller than one in model L and larger than one in models G1 and
G2. Therefore, any modulation in the observables at these early times
will strongly depend on the model considered. 
At later times, $t>4$~s, neutrinos with higher energies, $E\gsim 20$~MeV, 
feel the breakdown of the adiabaticity. At these energies, the
$\nuxbar$ flux is in all models larger than the  $\nuebar$ flux,  
$F^0_\nuxbar(E) > F^0_\nuebar(E)$. Therefore our prediction for the
observable modulation of SN neutrino spectra at later times, $t\gsim 4$~s,
will be model independent and thus more robust. For this reason, we
will basically concentrate the discussion on the modulation of the
observables arising at these times.

The main effect visible in the time evolution of $\la E_e \ra$ shown 
in the left panel of Fig.~\ref{time-an} is a decrease of $\la E_e \ra$ 
when the shock passes the resonance region and reduces the number of
events at high energies.  However, the details will depend on
the features of the shock propagation. If only the forward shock wave is
present, a deep and wide dip is expected. If moreover a reverse shock
is formed, a double dip will be observed: the positions of the two dips  
correspond to the times when one of the peaks in the survival
probability in Fig.~\ref{squarewave} coincides with the peak energy of
the antineutrino spectrum. 
In the case of $\xi=\la E_e^2 \ra / \la E_e \ra ^2$ the double dip structure
becomes a double bump. In contrast to an energy-independent survival
probability, $\xi$ is not always increased by mixing but can be also
reduced.  

It is remarkable that even in the model G2, where the original mean
energies $\la E_i \ra$ are exactly the same, the energy dependence of
the conversion probability results in an observable effect of the
shock wave propagation, because of the flavor-dependent fluxes.

Note that in this section we have assumed that the level crossings
are completely adiabatic except at the forward and reverse shocks.
For $\sin^2 \theta_{13}$ close to its current upper limit, this
is very likely to be the case. 
In the next section, we numerically calculate the survival 
probability without the foregoing assumption, and find that 
the qualitative features obtained here do indeed survive in
the complete numerical calculation. 
These features are thus robust, and though illustrated here
for the case of a special shock profile, should be
valid for a general case.

\section{Detection of the shock  and the reverse shock}
\label{numerical}

\begin{figure}
\epsfig{file=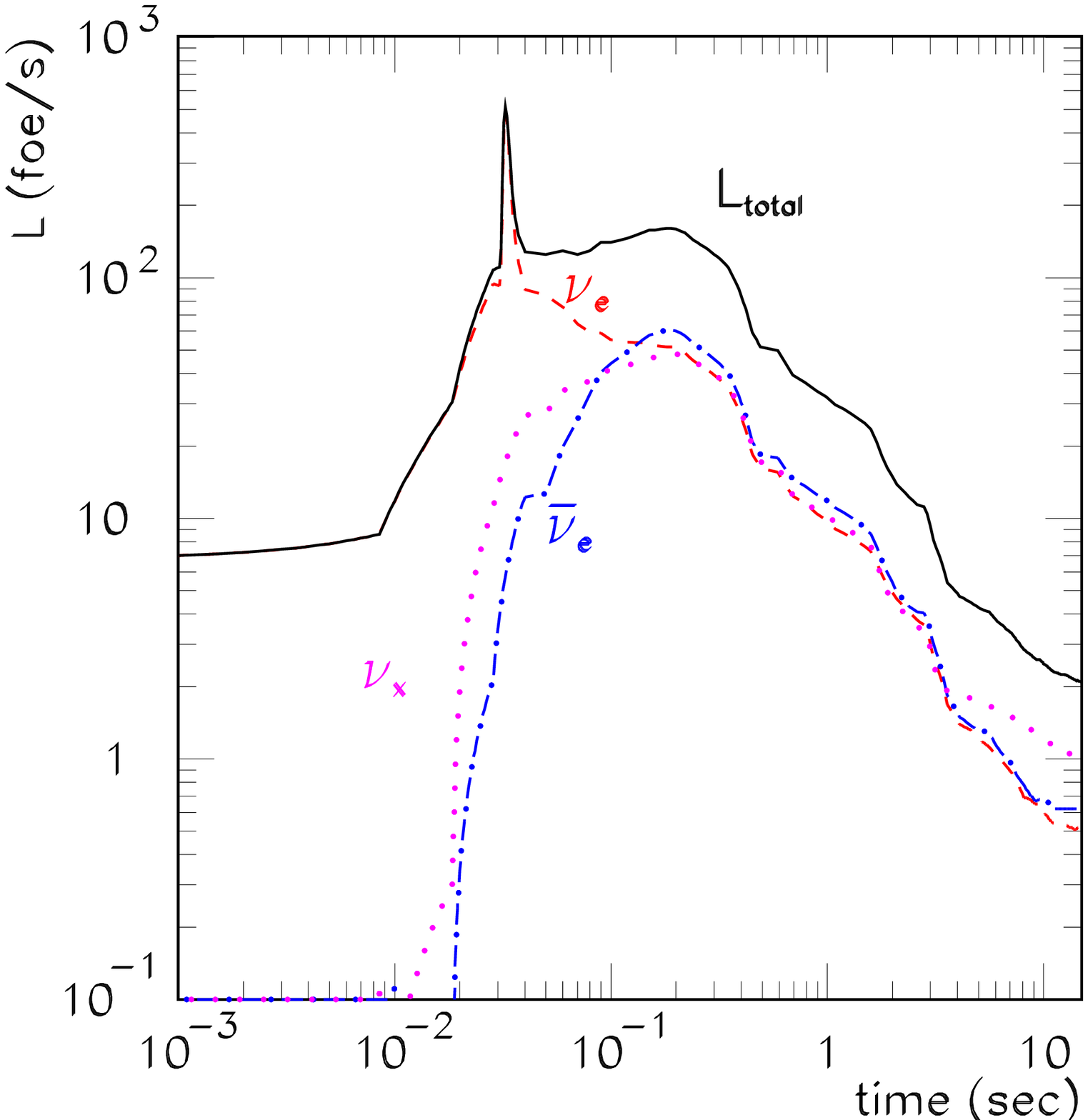,width=8cm}
\epsfig{file=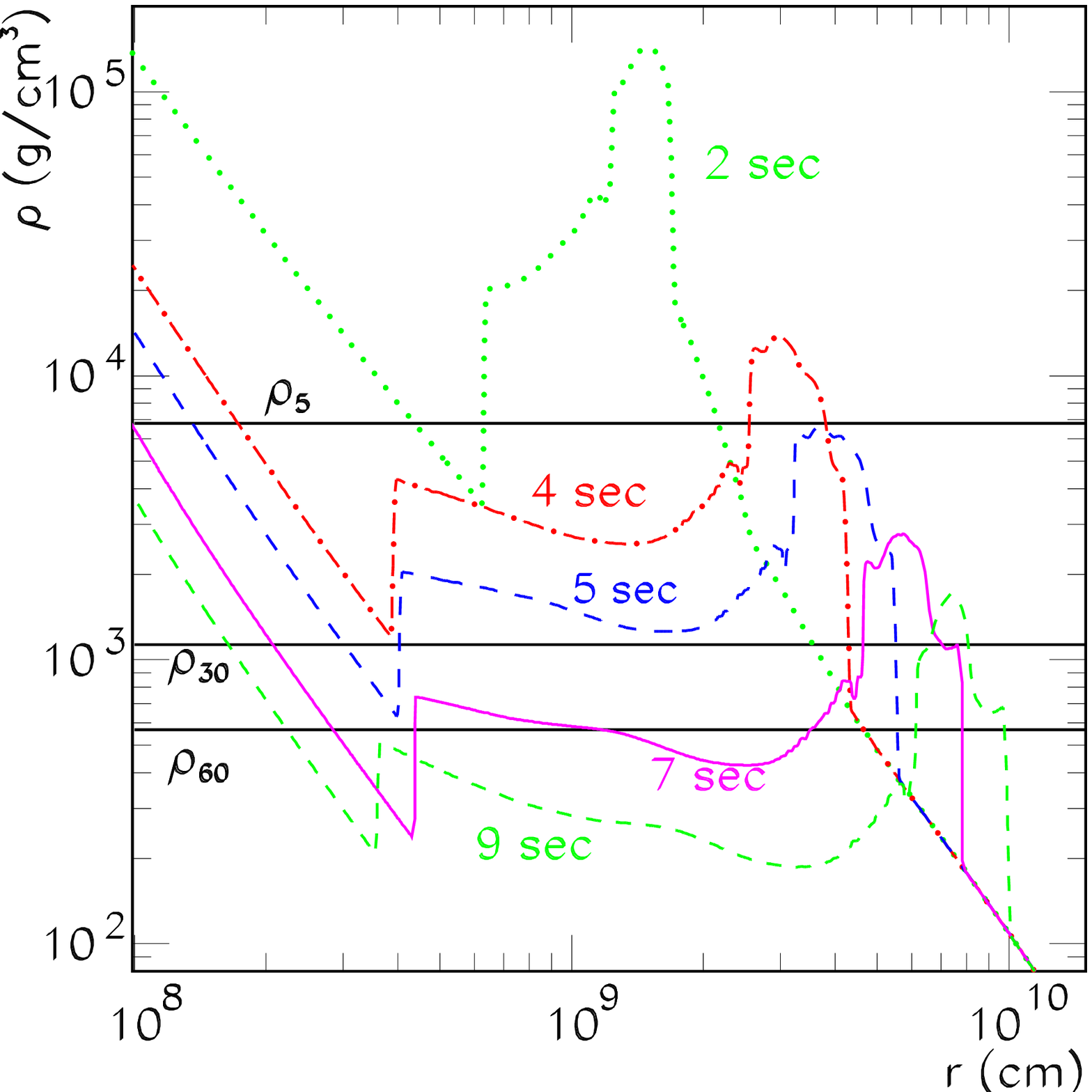,width=8cm}
\caption{Left: Time dependence of the neutrino luminosities $L$ used, from
  Ref.~\cite{livermore}. Right: Density profiles at different times 
  obtained from the numerical simulations of
  section 2. The H-resonance density, $\rho_{\rm H}$, for different neutrino
  energies, 5, 30 and 60 MeV, is also shown.
\label{numericalprofiles}}
\end{figure}

In this section, we examine in more detail the signatures of the forward
and reverse shock propagation imprinted on the neutrino signal of
a future galactic SN. The different ingredients used for the
simulation are

\begin{itemize}
\item
Initial fluxes: we assume for the time dependence of the total luminosity
the one given in Ref.~\cite{livermore}, shown in the left panel of  
Fig.~\ref{numericalprofiles}. We use the parameterization 
of Eq.~(\ref{flux-form}) for the spectral shape, with the values 
given in Tab.~\ref{tab:models}.
\item
The density profile of the SN envelope at different
times as discussed in Sec.~\ref{reverse}. In the right panel of
Fig.~\ref{numericalprofiles}, we show the effect of the forward and the
reverse-shock wave on the density profiles at several instances. 
The basic structures of the numerical profiles coincide well with the  
schematic profiles from Fig.~\ref{cartoon} used in our
analytical discussion, except at early times, where $\rho_{2a}\lsim
\rho_{1a}$. 
\item
The  numerically calculated survival probability $\bar p(E,t)$.
\item
Detector: we simulate the neutrino signal at a megaton water
Cherenkov detector assuming a SN distance of 10~kpc. The detector response is 
taken care of in the manner described in Ref.~\cite{pointing}.  The 
neutrino signal is dominated by the inverse beta reaction $\nuebar p 
\to n e^+$, while all other reactions have a negligible influence on the
analysis below. In future, the addition of Gd may allow for the efficient
tagging of this reaction~\cite{Gd}. Therefore, we take into account
only the inverse beta reactions in the following analysis.
\item
Neutrino mixing parameters: we concentrate on case B, i.e.\ an
inverted mass hierarchy and large $\theta_{13}$. In particular, we
use the following numerical values for the relevant parameters,
$\Delta m^2_\odot =  6.9\times 10^{-5}$~eV$^2$, 
$\tan^2\theta_\odot= 0.42$,
$\Delta m^2_{\rm atm}=2.6\times 10^{-3}$~eV$^2$, and
$\tan^2\theta_{13}=10^{-2}$.
However, we have checked that the signatures of the shock propagation
like the ``double-dip'' feature are detectable even for
$\tan^2\theta_{13}=5\times 10^{-5}$. 
\end{itemize}

\begin{figure}
\epsfig{file=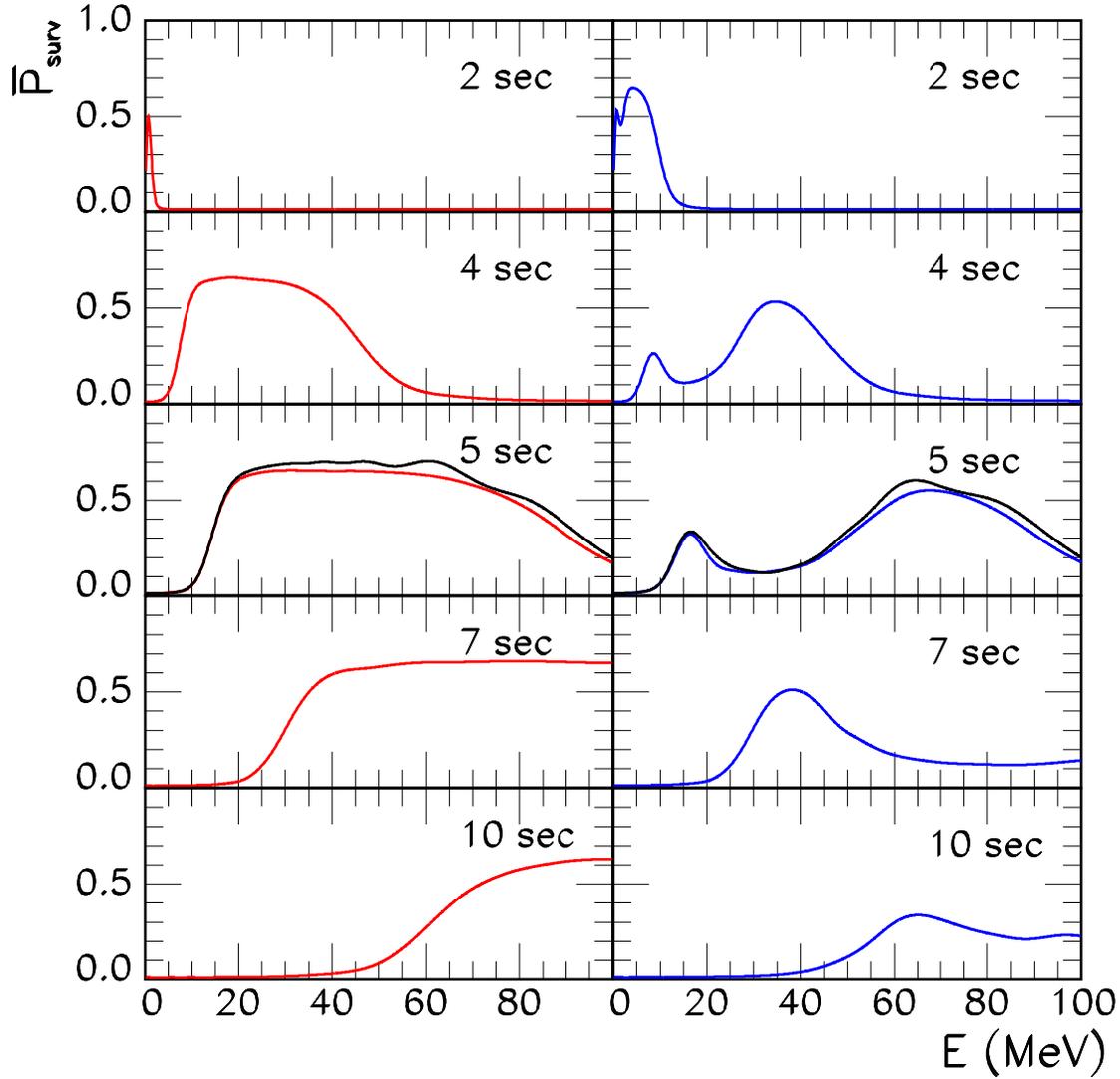,width=16cm}
\caption{Survival probability $\bar p(E,t)$ as function of energy at
  different times averaging in energies with the energy resolution of
  Super-Kamiokande; for a profile with only a forward shock (left) and a
  profile with forward and reverse shock (right).
  At $t=5$~s, we show $\bar p(E,t)$ including Earth matter effects for a zenith
  angle of the SN of $62^\circ$ (black line).
\label{pee}}
\end{figure}

The key ingredient to observe signatures of the shock wave propagation is
the time and energy dependence of the neutrino survival probability.
In Fig.~\ref{pee}, we show the energy dependence of $\bar p(E,t)$ averaged
with the energy resolution function of Super-Kamiokande, for the case with
(right panel) and without (left panel) a reverse shock. To simulate
the latter case, we removed by hand the reverse shock from the
numerical density profiles.

Qualitatively, we observe the same features as using the analytical
approximation of Sec.~\ref{analytical}: 
While the survival probability  has a single peak if there is only a
forward shock propagating, the reverse shock superimposes a dip at
those energies for which the resonance region is passed by both shock
waves. All these structures move in time towards higher energies, as
the shock waves reach regions with lower density.
The width and location of the peaks depend on the particular values of
$E_{1a},E_{1b},E_{2a},E_{2b}$ at each instant. We can observe, for
example, that at $t=2$~s only low energy neutrinos, $E\sim 10$~MeV,
are affected by the reverse shock. A little bit later, at $t=5$~s, 
neutrinos with energies $E\sim 10$--15~MeV are influenced by
the forward shock wave (the first peak), while those with $E\sim 20$--60~MeV
feel the effect of both the reverse and the forward shocks. The 
effect of the two strongly non-adiabatic transitions cancel partially
and $\bar p(E,t)$ has a valley around $\sim 30$~MeV. For $E\gsim
60$~MeV, neutrinos cross only one shock wave, thus a second peak
appears in $\bar p(E,t)$. At later times, this pattern is basically
repeated, only shifted to higher and higher energies.

\begin{figure}
\vskip-0.3cm
\epsfig{file=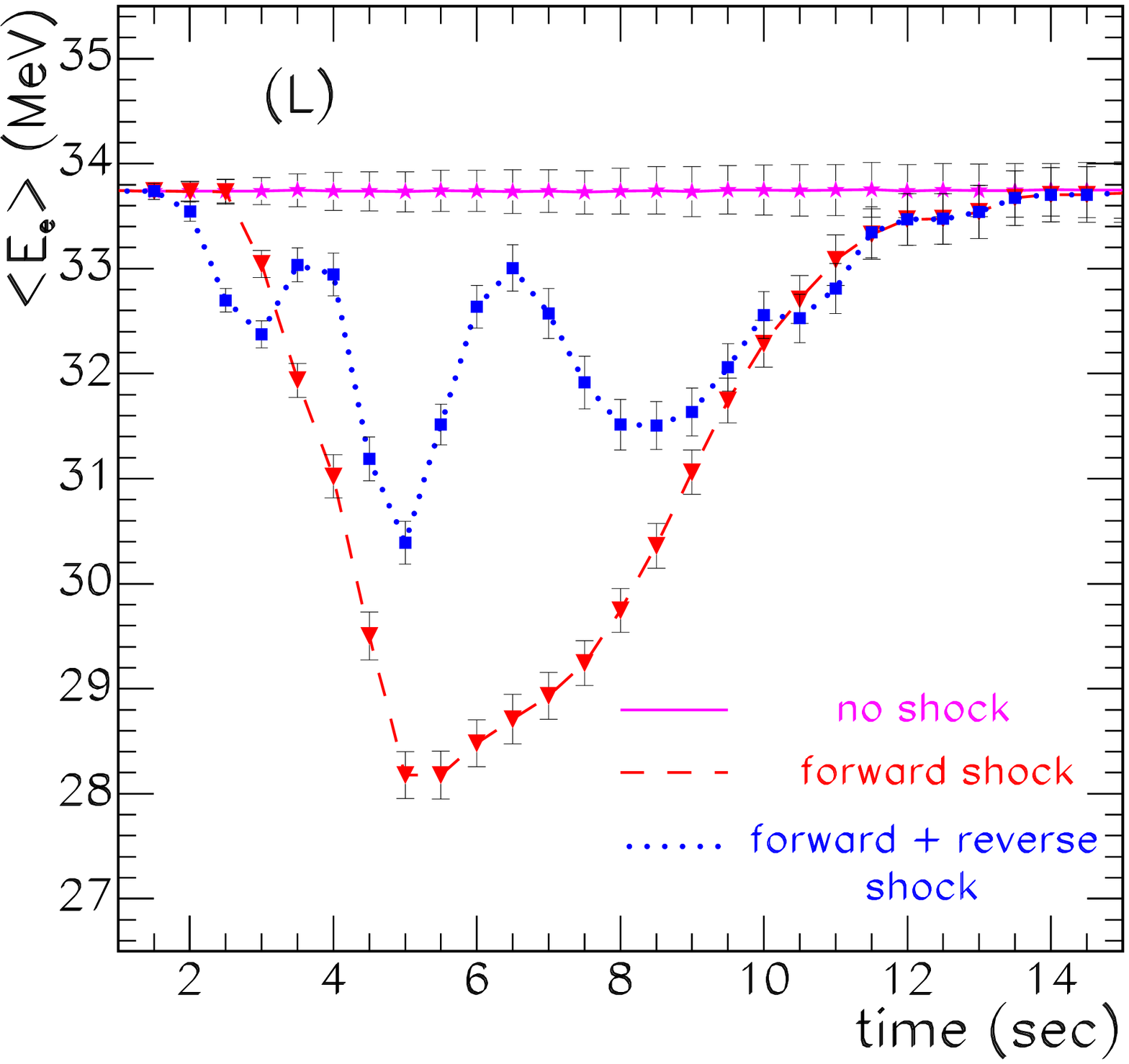,width=8cm,height=7.5cm}
\epsfig{file=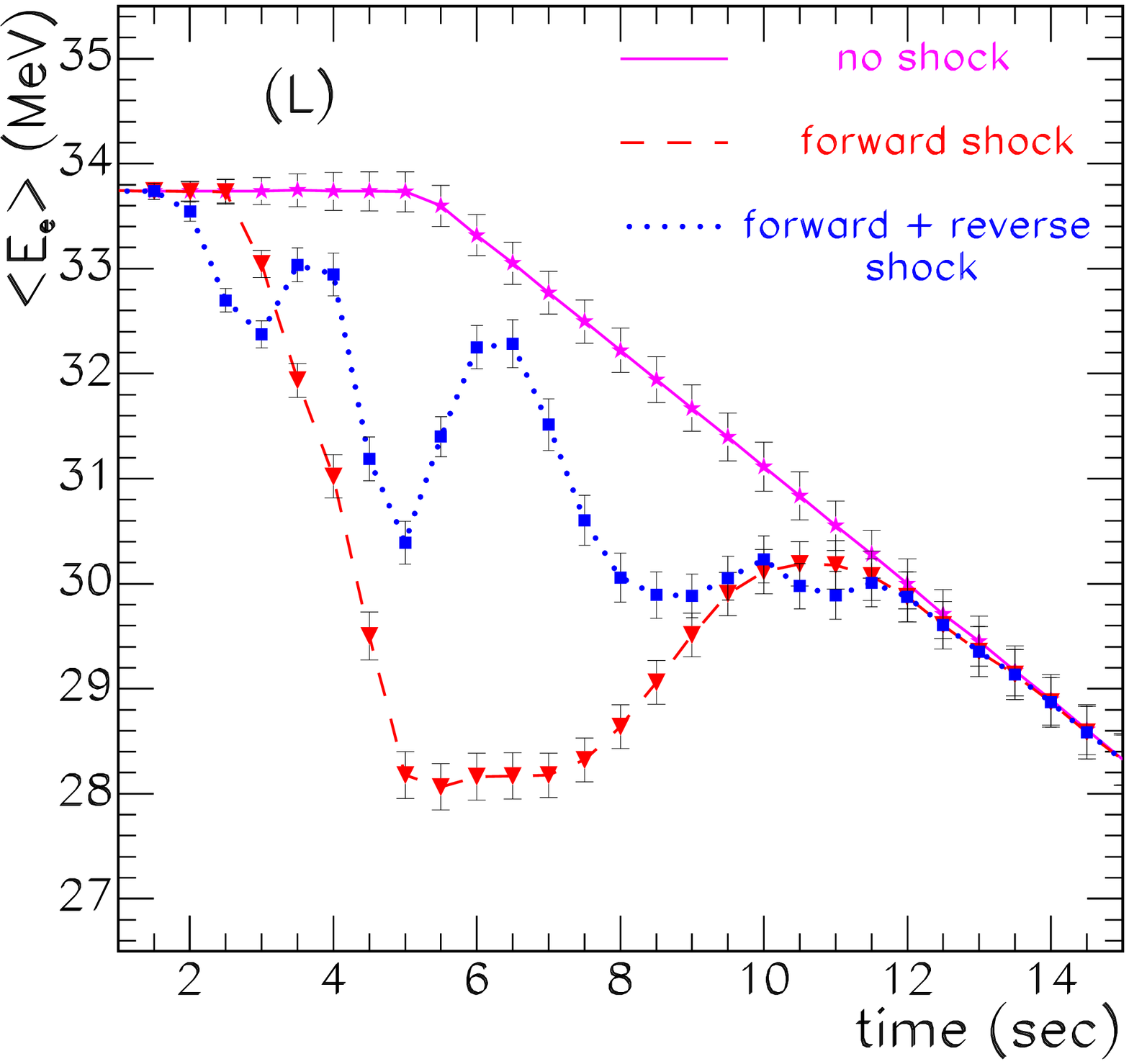,width=8cm,height=7.5cm}
\vskip-0.5cm
\epsfig{file=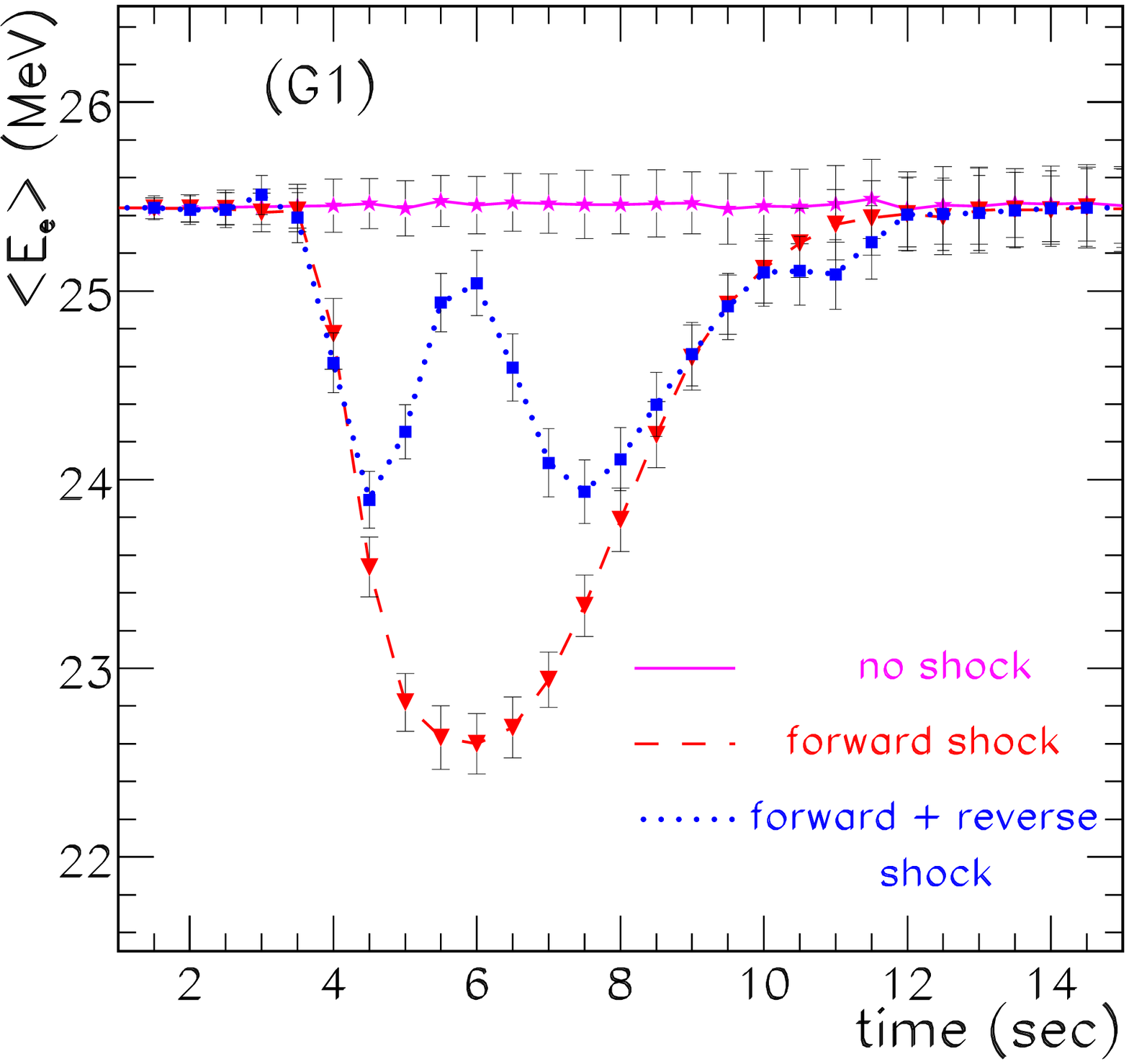,width=8cm,height=7.5cm}
\epsfig{file=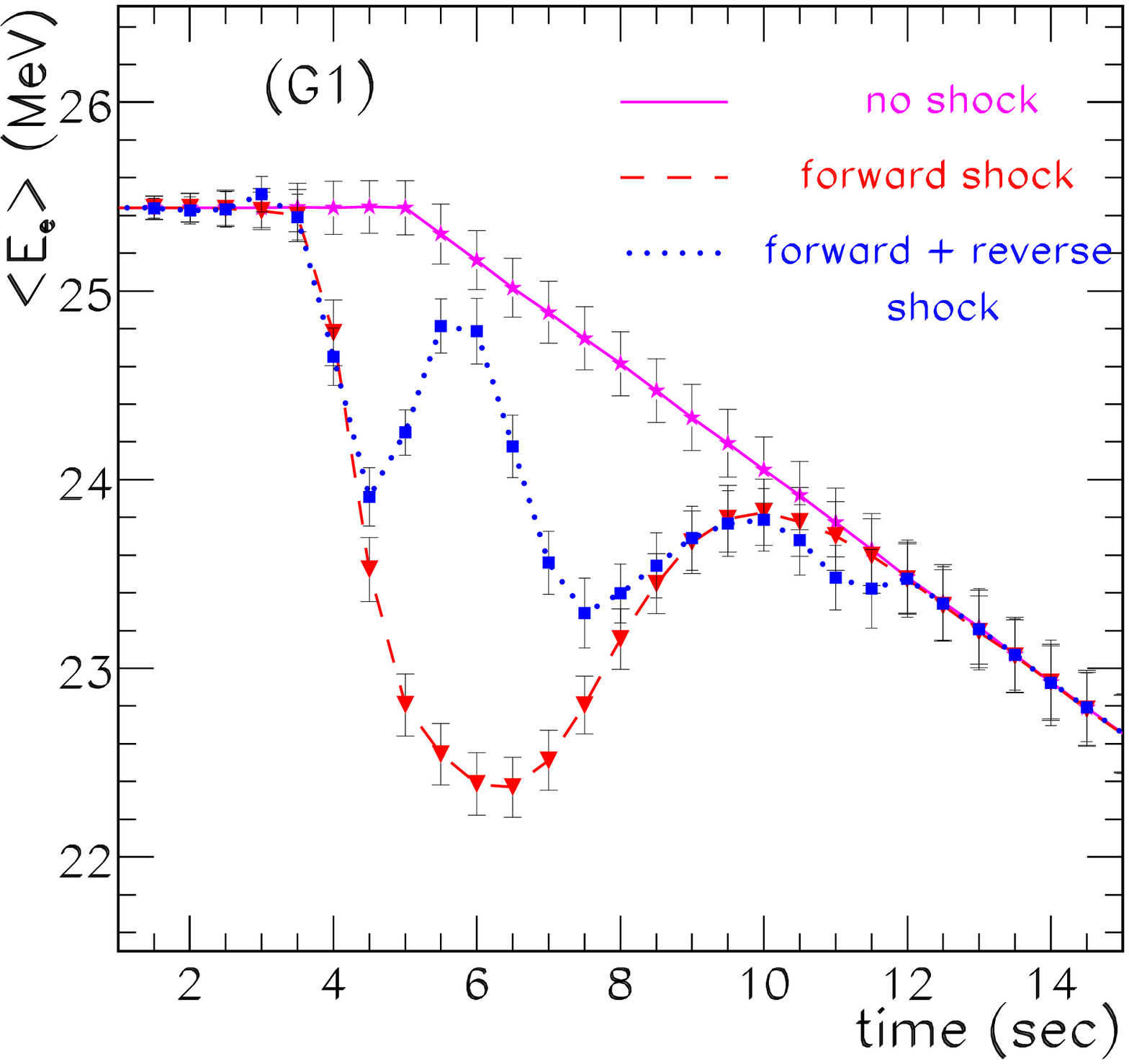,width=8cm,height=7.5cm}
\vskip-0.5cm
\epsfig{file=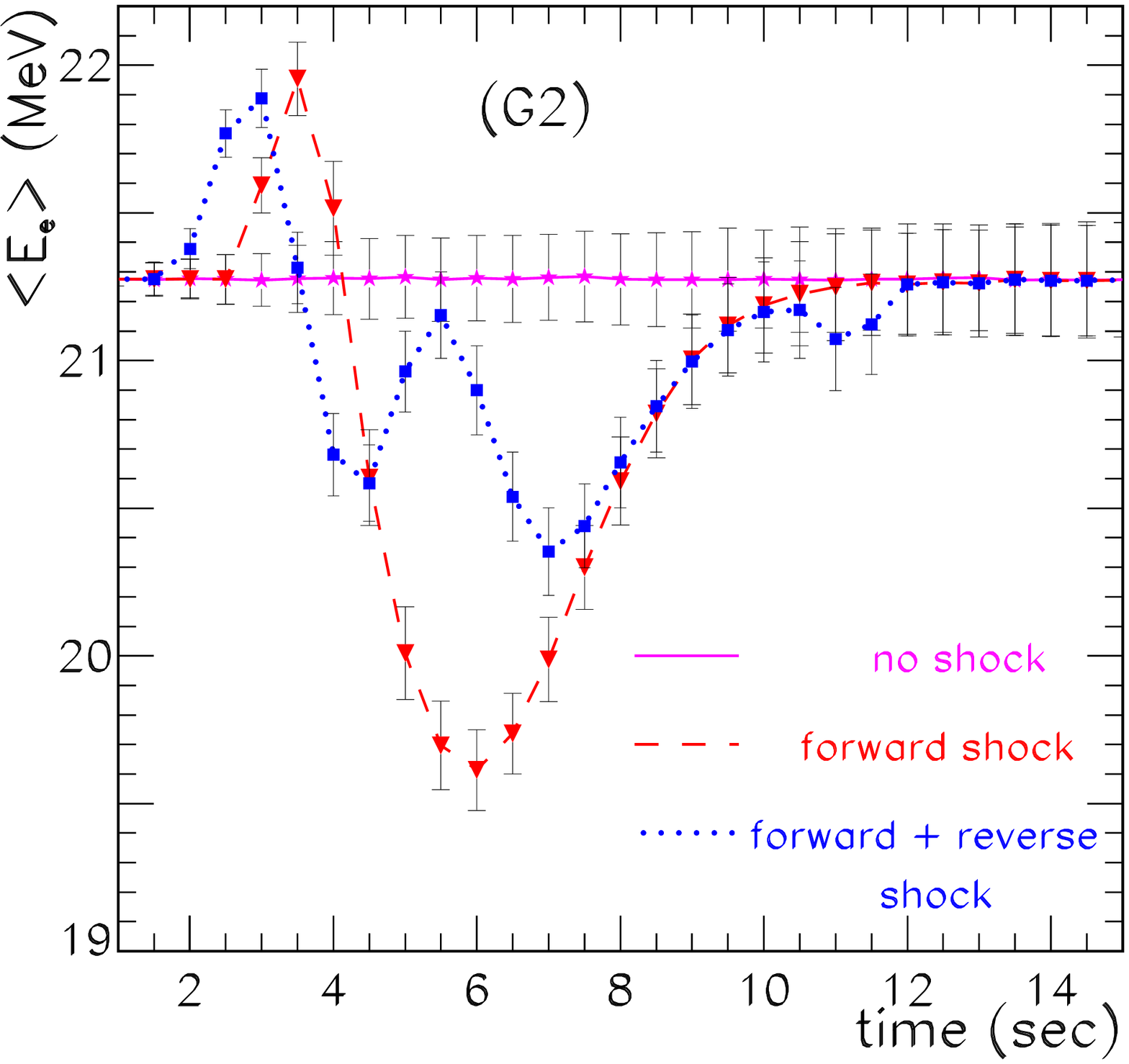,width=8cm,height=7.5cm}
\epsfig{file=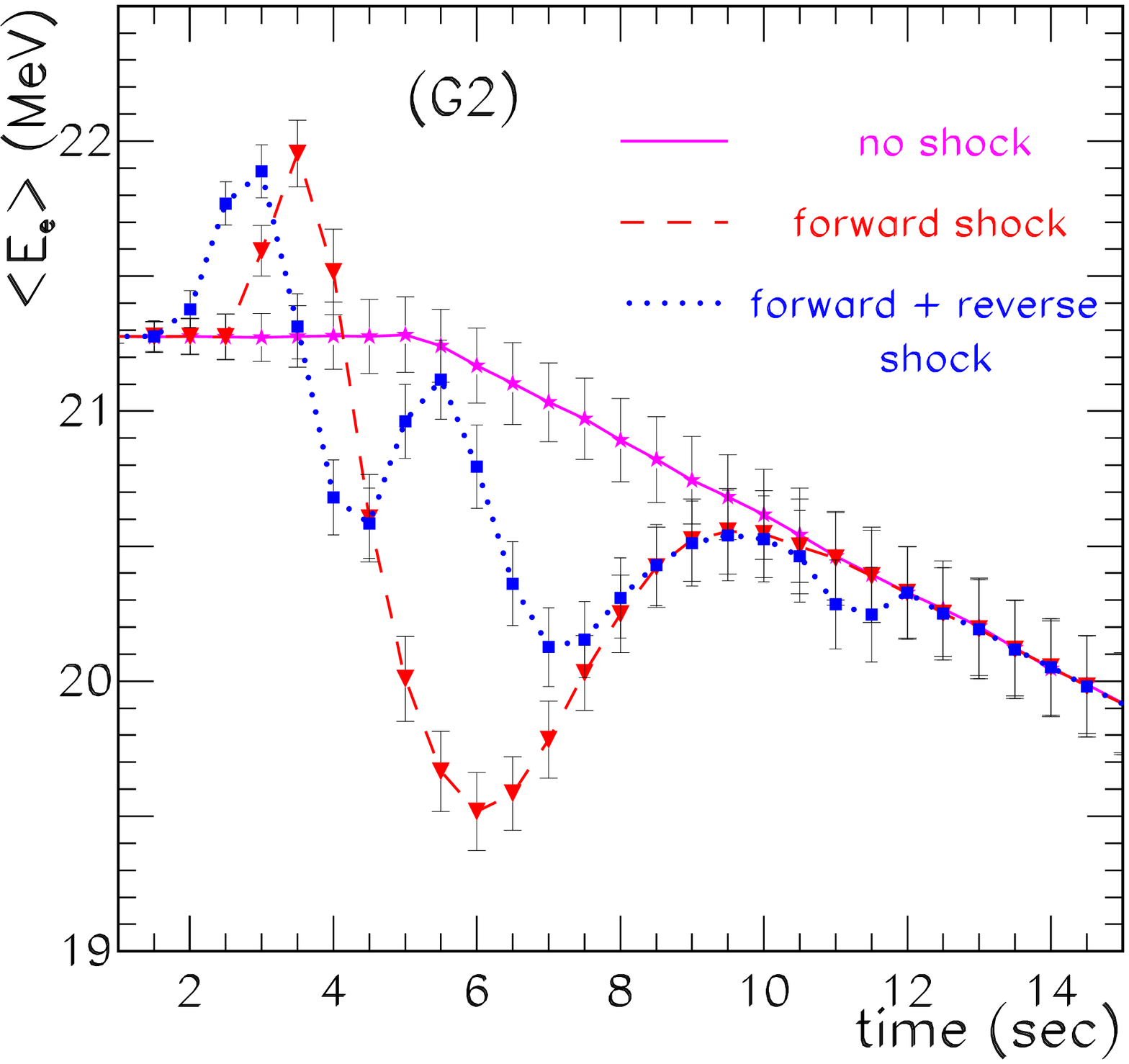,width=8cm,height=7.5cm}
\caption{The average energy of $\bar\nu p\to ne^+$ events binned in time 
for a static density profile (magenta), a profile with only a forward
shock (red) and with forward and reverse shock (blue). In the left
panels we use $\la E_i \ra={\rm const.}$, whereas in the right
panels $\la E_i \ra$ decreases for $t>5$~s.
The error bars represent 1~$\sigma$ errors in any bin.
\label{Eav}}
\end{figure}

The first observable we consider is the average of the measured
positron energies, $\la E_e\ra$, with a time binning of 0.5~s. In
Fig.~\ref{Eav}, we show $\la E_e \ra$ for various cases
together with the one sigma errors:
Each of the six panels contains the case that no shock wave
influences the neutrino propagation, the case of only a forward shock
wave and of both forward and reverse shock wave. In the three left panels
the $\la E_i \ra$ are constant in time, while in the three right panels
we assume that the $\la E_i \ra$ decrease linearly after 5~s. Finally, 
the upper, middle and lower panels show the three models L, G1 and
G2 for the primary neutrino spectra given in Table~\ref{tab:models}. 

\begin{figure}
\epsfig{file=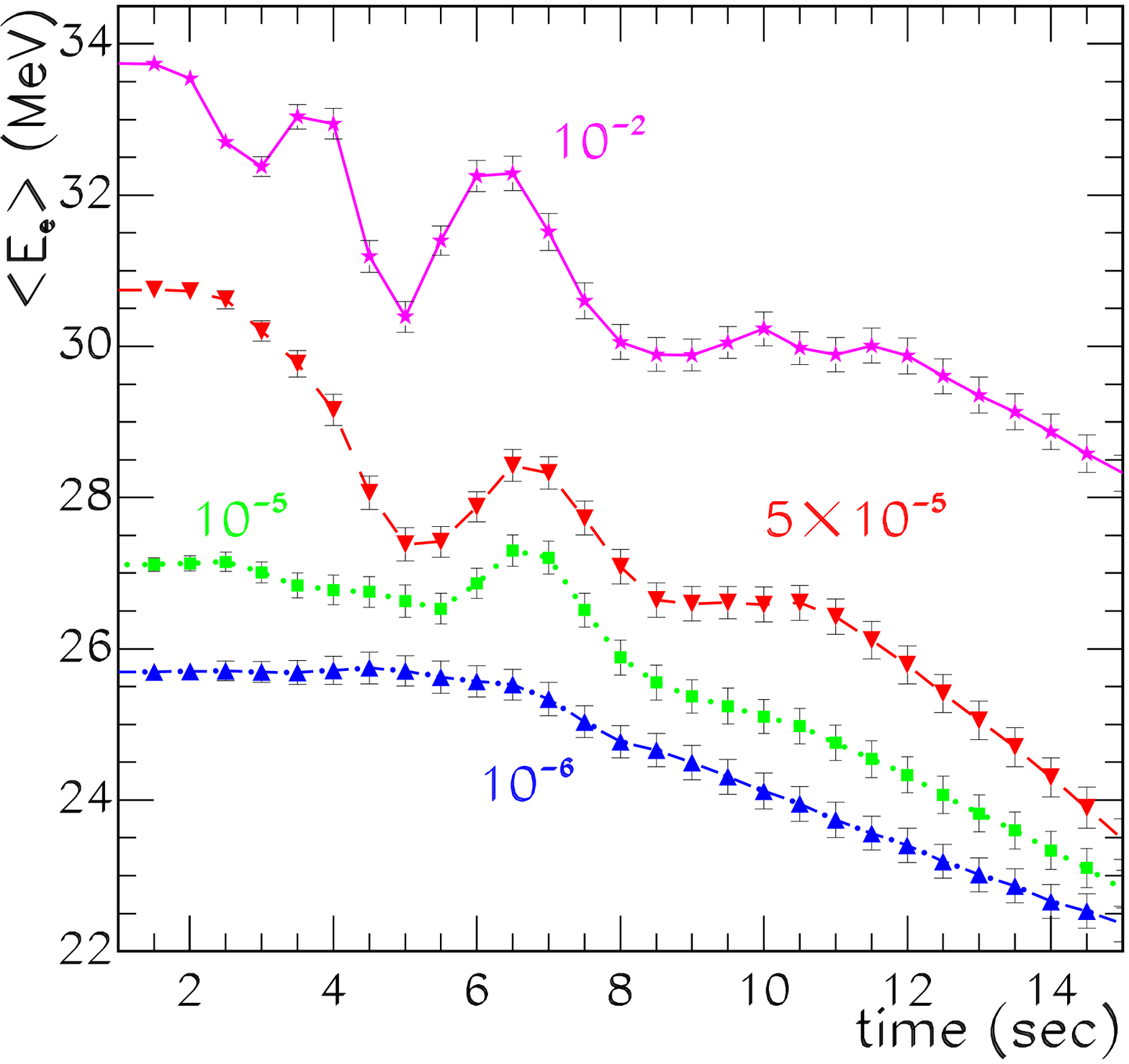,width=8cm}
\epsfig{file=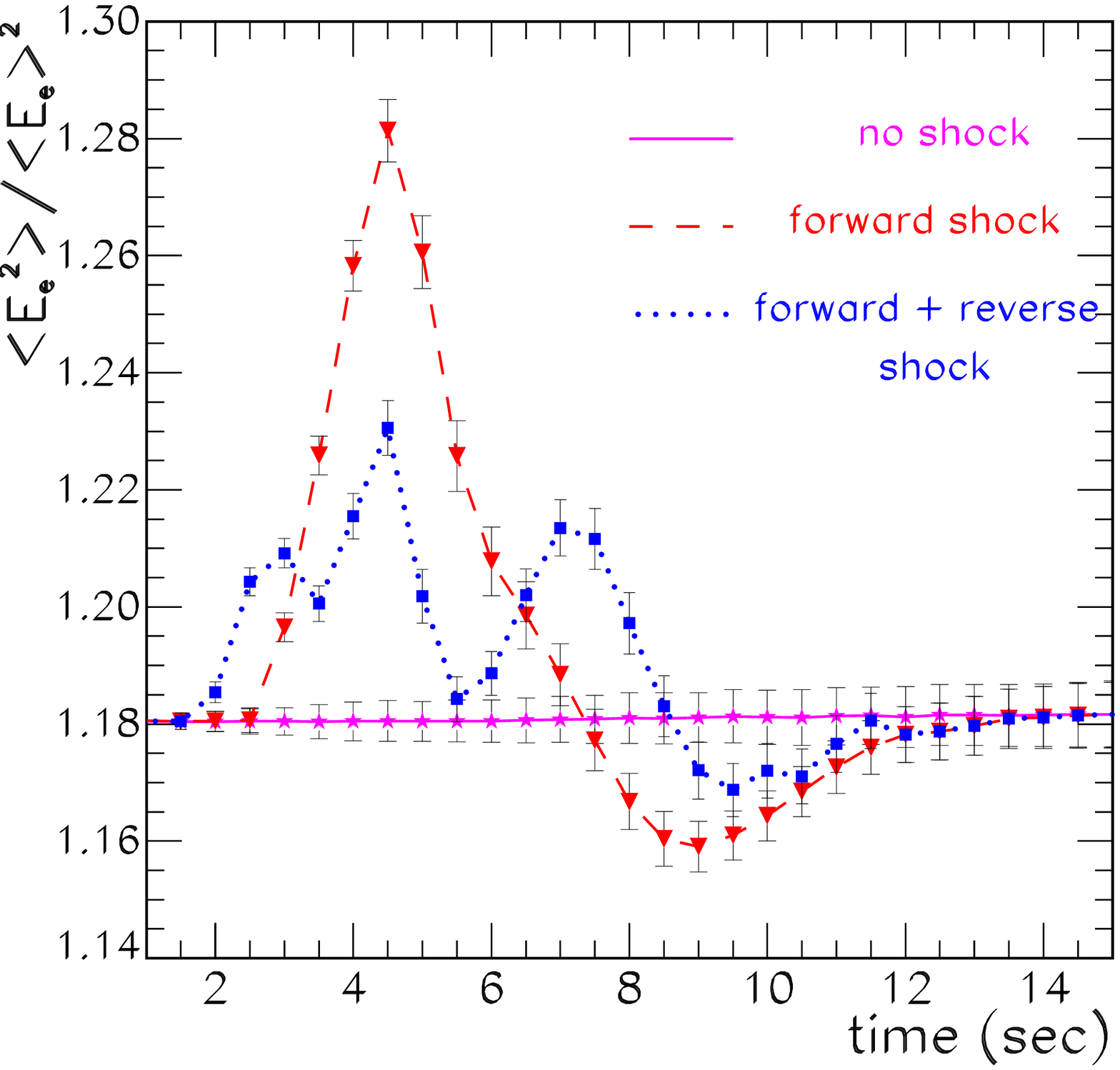,width=8cm}
\caption{
  Left: Time dependence of $\la E_e\ra$ 
  for a profile with a forward and reverse shock for
  several values of $\tan^2\theta_{13}$ as indicated; for model L.
  Right:
  Time dependences of the observable $\xi=\la
  E_e^2 \ra / \la E_e \ra^2$ 
  for a static density profile (magenta), a profile with only a forward
  shock (red) and with forward and reverse shock (blue). 
\label{alpha}}
\end{figure}

As our most important result we find that the effects of the shock wave
propagation are clearly visible in the neutrino signal of a megaton
detector, independent of the assumptions about the initial neutrino
spectra. Moreover, it is not only possible to detect the shock wave
propagation in general, but also to identify the specific 
imprints of the forward and reverse shock versus the forward shock only  
case. The signature of the reverse shock is its double-dip structure
compared to the one-dip of a forward shock only,
as predicted by our analytical model in Fig.~\ref{time-an}. 
While the presence or absence of structures at early times does depend
on the initial fluxes and the details of the density profiles, the
two-dip pattern is a robust signature of the presence of a
reverse shock. 
To study the dependence of the double-dip structure on the value of
$\theta_{13}$, we show $\la E_e\ra$ as function of time for different
13-mixing angles in the left panel of Fig.~\ref{alpha}. Even for as
small values as $\tan^2\theta_{13}=5\times 10^{-5}$ the double-dip is
still clearly visible, while for $\tan^2\theta_{13}=1\times 10^{-5}$
only a bump modulates the neutrino signal.

Next we discuss the possibility to detect the imprint of the shock
waves in other observables. To shorten the exposition, we consider
only one model (L) together with a linear decrease of $\la E_i\ra$
after $t=5$~s.
In Fig.~\ref{alpha}, we show the time dependence of the observable
$\xi = \langle E_e^2 \rangle / \langle E_e \rangle^2$.
If the shock does not influence the neutrino spectra,
i.e. $\bar{p}\approx 0$, then $\xi = \xi_0 =
(\beta_\nuxbar+ 3)/(\beta_\nuxbar+ 2)$.
For a mixed spectrum, with $\bar{p}(E)$ independent
of energy, we have $\xi > \xi_0$ in general.
However, for a strongly energy dependent $\bar{p}(E)$
as in the case of a shock, $\xi$ can increase or decrease
depending on the details of $\bar{p}(E)$ and the
primary spectra.
Such modulations in the value of $\xi$ that coincide in time
with the modulations in $\langle E_e \rangle$
act as a confirming evidence for the passage
of the shock wave.

\begin{figure}
\epsfig{file=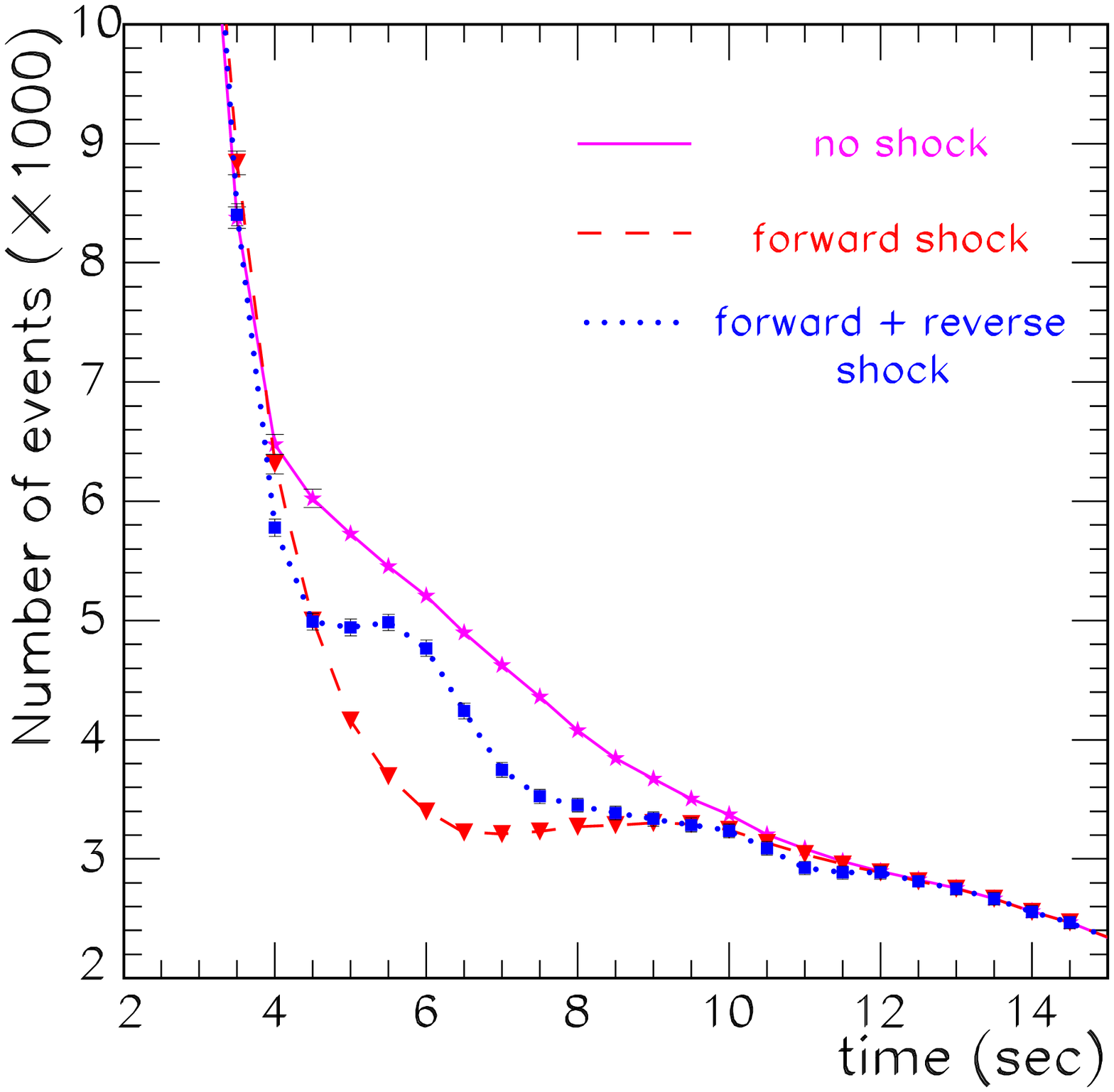,width=8cm}
\epsfig{file=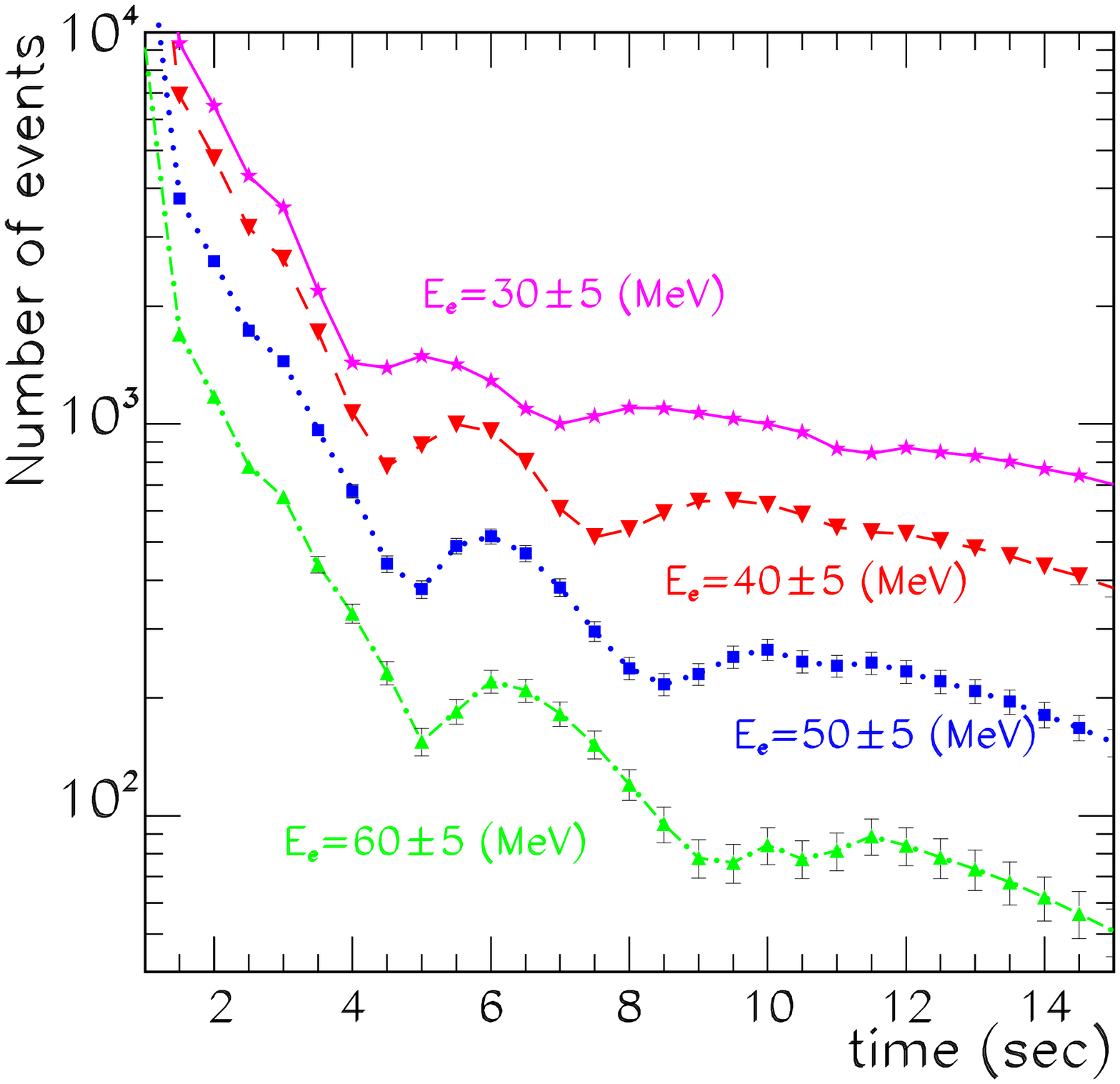,width=8cm}
\caption{Left: Total number of events detected as function of time. 
  Right:  Number of events binned per energy decade as function
  of time for forward and reverse shock. 
\label{N}}
\end{figure}

Another observable that can be exploited is the total number of
detected events. In the left panel of Fig.~\ref{N}, we show the number
of events binned in time. This observable is particularly interesting
for detectors like IceCube with poor or no energy resolution at all
for SN neutrinos.
If the luminosities are fast decreasing with time, the modulations
introduced by the time-behavior of $\bar p(E,t)$ may be difficult to
disentangle from the overall time-behavior of the luminosities.
Nevertheless, even a detector without energy resolution 
like IceCube has a potential to discover the imprint of
propagating shock waves on the neutrino signal, if the luminosities are
not decreasing as fast as found in Ref.~\cite{livermore}.

Since the information about the shock fronts is encoded in the energy
and time dependence of the survival probability $\bar p(E,t)$, one can
expect that the number of events in a fixed energy range is a more useful
observable than the total number of events. In the right panel of
Fig.~\ref{N},  we show therefore the number of events binned
in energy intervals of 10~MeV as function of time for the case of a
reverse shock. We can observe clearly how the positions of the two dips
change in each energy bin. It is remarkable that the double-dip
feature is not only stronger after binning, but also allows one to trace 
the shock propagation: Given the neutrino mixing scheme, the neutrino
energy fixes the resonance density. Therefore, the progress of the
shock fronts can be read off from the position of the double-dip in
the neutrino spectra of different energy.

\begin{figure}
\epsfig{file=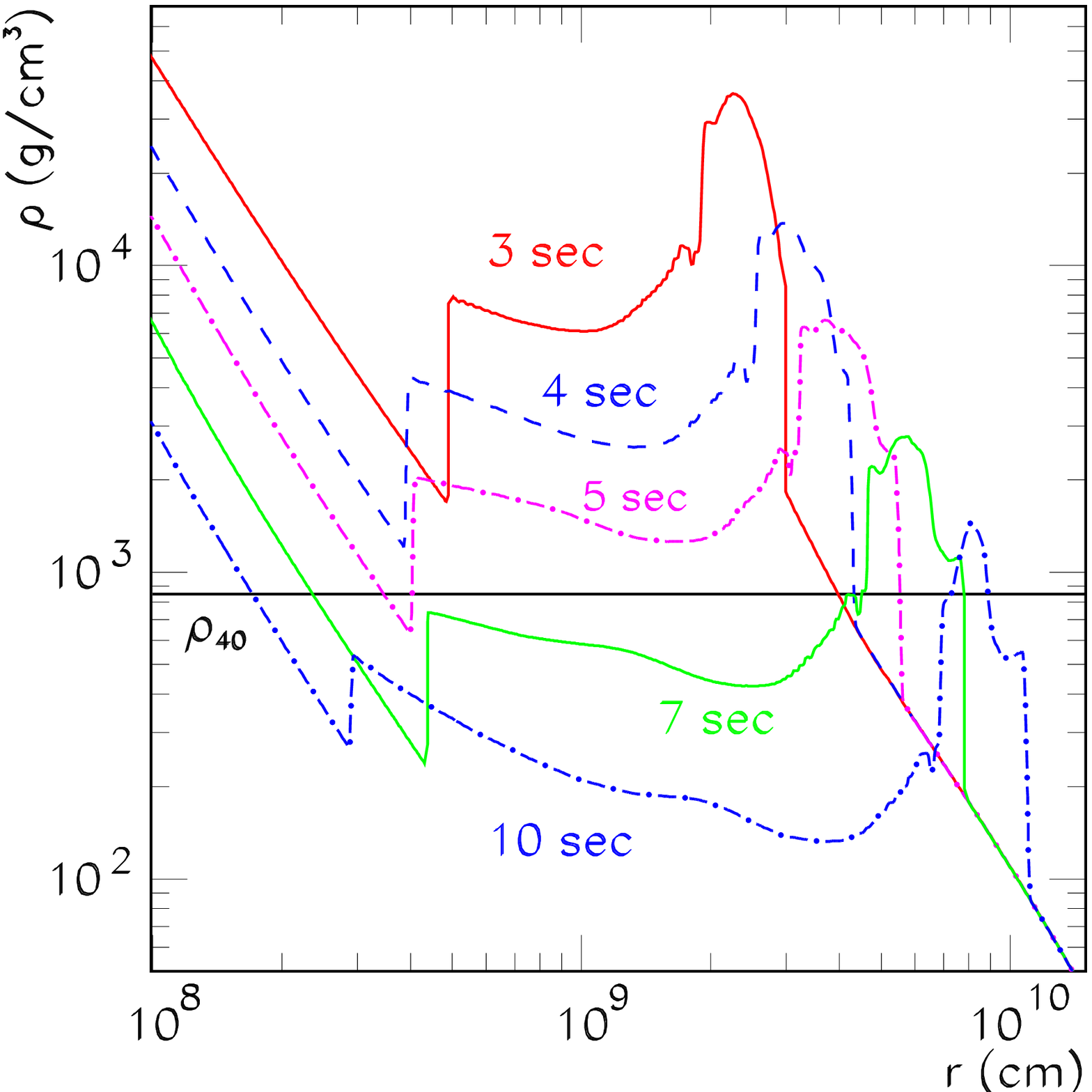,width=8cm}
\epsfig{file=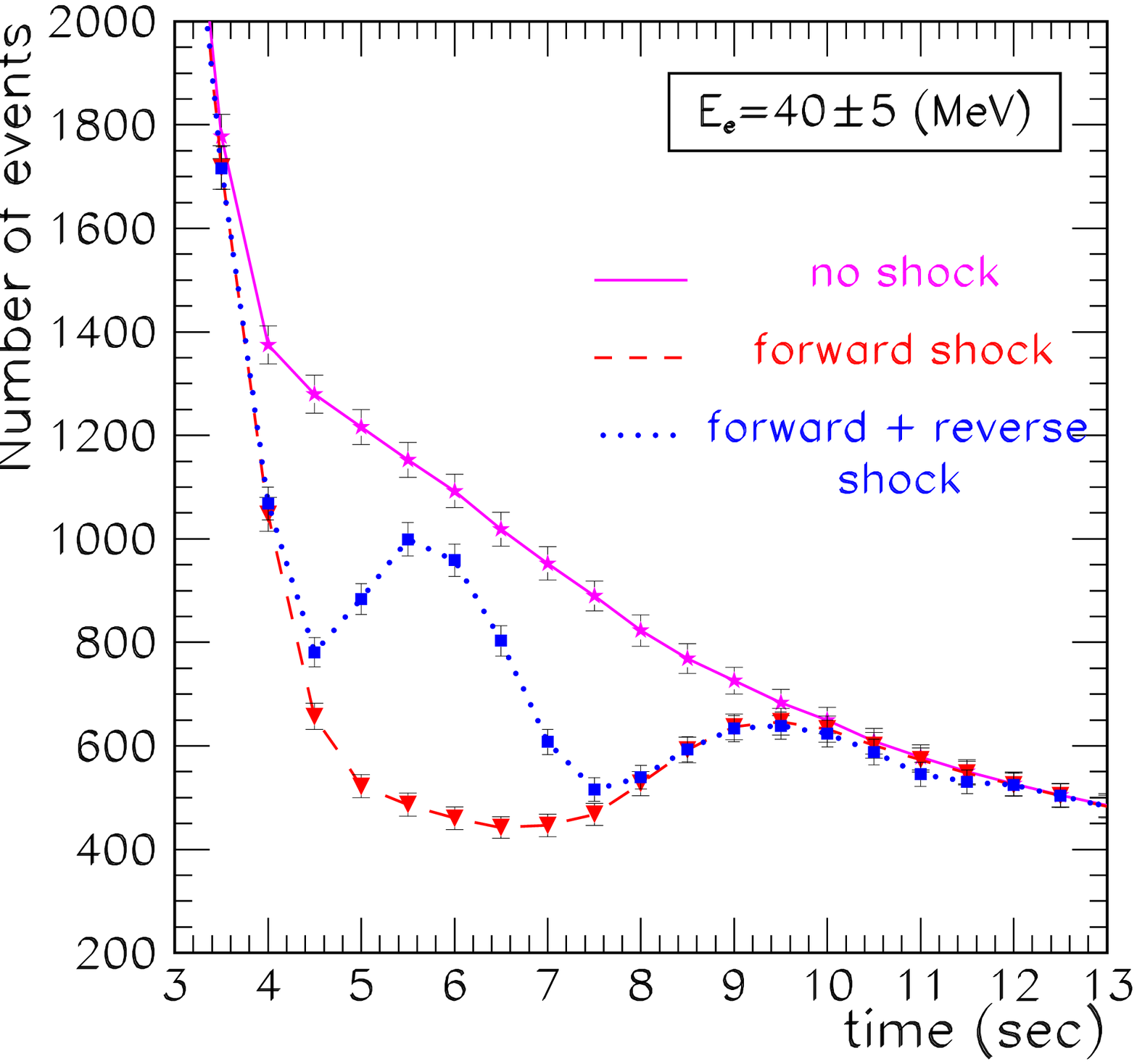,width=8cm}
\caption{
Left: Density profiles at different times with
  the resonance layer of neutrino with $E=40$ MeV superimposed.
Right: Number of events expected in the energy range $E=40\pm
  5$~MeV as function of time. 
\label{Ebinned}}
\end{figure}

We illustrate this by examining the time evolution of the number of
events in the bin $40 \pm 5$~MeV in detail. From Fig.~\ref{Ebinned},
right panel, it can be read off that between 4.5--7.5~s, the neutrinos
with this energy pass through two nonadiabatic resonances, i.e.
both the forward and the reverse shock are present in
the regions with density $\rho_{40} \approx 850$~g/cm$^3$.
Between 7.5~s and 9.5~s, only one of the shock fronts is
present at this density. This inferred behavior of the shock wave can
be seen to correspond with the shock profile used, see the left panel
of Fig.~\ref{Ebinned}. For times before 4.5~s and after 9.5~s,
the data is not able to say anything concrete about the shock wave
propagation.

If only the forward shock is present, then Fig.~\ref{Ebinned}  
still allows us to
infer that one shock wave was present in the region $\rho_{40}$ between
5--9.5~s. No concrete conclusions can be drawn about the behavior
of the shock wave beyond this time interval.

The number of shock waves present in a region with any given density 
$\sim 10^3$~g/cm$^3$ can
similarly be extracted from the data by considering the time evolution of
the number of events in the energy bin corresponding to that density.
An extrapolation would allow one to trace the positions of the forward
and the reverse shock waves for times between 1--10~s.

\begin{figure}[t]
\epsfig{file=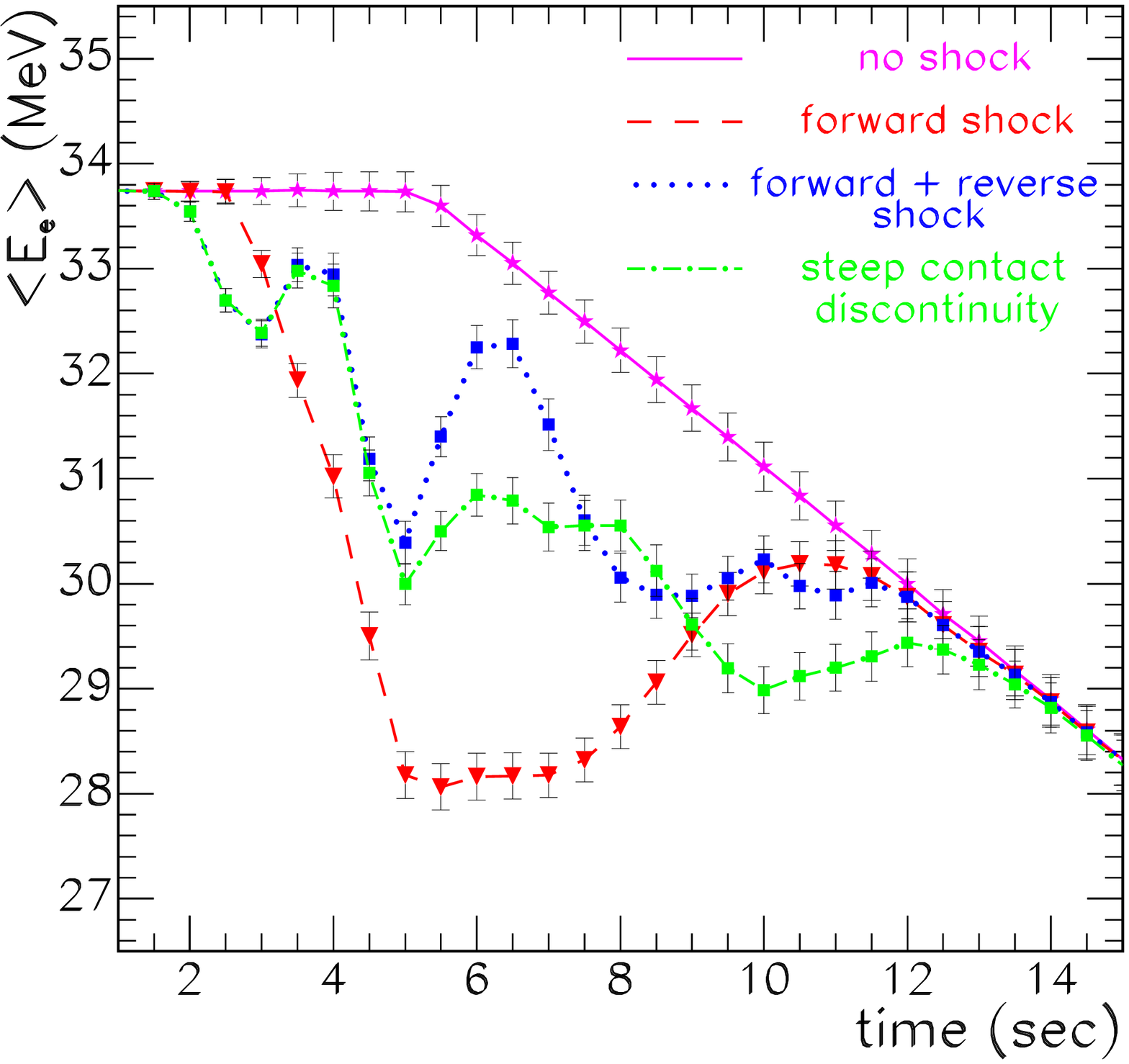,width=8cm}
\epsfig{file=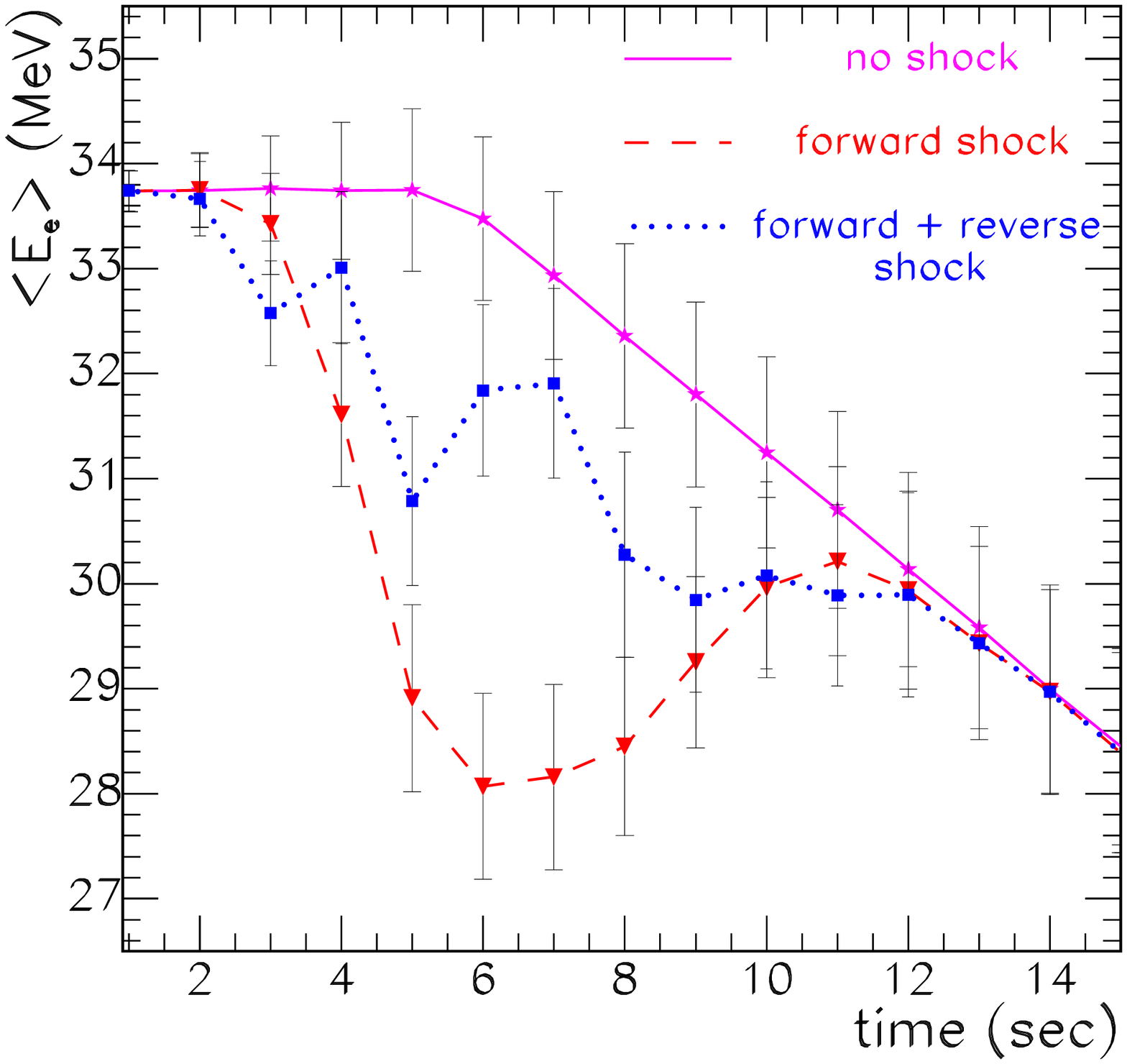,width=8cm}
\caption{
  Left: Observed $\la E_e \ra$ for a static density profile
  (magenta), a profile with only a forward shock (red), with forward and
  reverse shock (blue), and in the case that the contact discontinuity
  is steepened by hand (green). 
  Right:
  Observed $\la E_e \ra$ in Super-Kamiokande, color code as in the
  left panel.
\label{discontinuity}}
\end{figure}

Until now we have neglected the possibility that other discontinuities
in addition to the forward and reverse shock can influence the neutrino
propagation. However, because of the decreasing resolution in our
simulation at large radii as described in Sec.~2, we are only able to
resolve the outward moving contact discontinuity for 2--3 seconds. Its 
width at that time is found to be 200--300$\,$km in our one-dimensional 
model. Hydrodynamic instabilities are expected to widen the contact
discontinuity, and we 
expect therefore that the neutrino evolution is adiabatic also at
later times at this discontinuity.
But to get an idea of the possible influence of this discontinuity on
observables, we have steepened it by hand for
$t>3$~s, so that the neutrino evolution becomes strongly
non-adiabatic at this point. 
In Fig.~\ref{discontinuity}, we show the detected $\la E_e \ra$ in this case.
Since now there are three shocks (and six associated energies), the
structures in $\la E_e \ra$ become more complicated but the imprint of
the shock wave is still clearly visible. In particular, there are
effects of the shock wave propagation at later times than in the
previous cases, because for the density profiles considered in our
study the largest density affected by this new discontinuity is larger
than the maximum of $\rho_{1a}$ and $\rho_{1b}$. 

Finally, we want to discuss if the largest currently operating detector, 
Super-Kamiokande, can observe the signatures of shock wave propagation.
Compared to a megaton detector, the event rate in Super-Kamiokande is
reduced by a factor of 30. Therefore, we plot in the right panel of
Fig.~\ref{discontinuity} the mean event energy $\la E_e \ra$   
with one second time bins in order to decrease the statistical error. 
The features of the different shock propagation scenarios remain
visible, but are now more difficult to disentangle because of the
larger errors. However, one should keep in mind that the number of
events is not only related to the volume of the detector but also to 
the distance to the SN. Thus, the potential of
Super-Kamiokande to yield important clues about the shock propagation
cannot be dismissed.

\section{Summary}
\label{concl}

We have performed simulations of supernova explosions adjusting the
neutrino energy transfer to the shock so that a successful 
explosion resulted. Following the time evolution  in spherical symmetry 
until more than 25~seconds after shock formation, we have found that
in addition to the forward shock wave a reverse shock forms when the
supersonically expanding neutrino-driven wind collides with the slower
earlier SN ejecta. Both the forward and reverse shock are sharp
discontinuities where the density changes on a sub-millimeter scale. 

The sudden density jump at the two shock fronts results in a strongly 
non-adiabatic evolution of neutrino flavor oscillations, when the shocks 
cross the H-resonance layer. For a ``large'' 13-mixing angle,
$\sin^2\theta_{13}\gg 10^{-5}$, the MSW enhancement of flavor
conversion that is otherwise working is interrupted during the shock
passage. This break-down of adiabaticity results in a reduction of the
average energy and number of detected $\bar\nu_e$ events for an
inverse neutrino mass hierarchy (case B) or of $\nu_e$ events for a
normal hierarchy (case A). The reduced event rate during the
shock passage through the H-resonance allows even experiments with
missing energy resolution for SN neutrinos, as e.g. IceCube, to observe
effects of the shock propagation. 

The characteristic signature for the presence of two shocks is the
``double-dip'' feature  in the time-binned energy spectrum of
observed electron-neutrino events. We have found that already
Super-Kamiokande can potentially distinguish between
different shock propagation scenarios, while a megaton detector
may be able to trace the time-evolution of forward and reverse shocks.

The modulation of SN neutrino spectra by propagating shocks is only
possible if $\theta_{13}$ is ``large,'' $\sin^2\theta_{13}\gg 10^{-5}$. 
Observing these features in the neutrino signal of a future galactic
SN would therefore indicate that the measurement of $\theta_{13}$
and the detection of leptonic CP violation may be within reach.
In addition, the detection of any modulation in the $\bar\nu_e$ or
the  $\nu_e$ spectrum by shock effects would identify the neutrino
mass hierarchy.  
Remarkably, observing features of SN shock propagation is
complementary to the detection of Earth matter effects: while observing
modulations by SN shocks in the $\bar\nu_e$ spectrum identifies
case B, modulations by Earth matter effects exclude this case. 

We want to close with a speculative remark about the time-structure of
the neutrino events from SN~1987A. The events observed by the Kamioka
experiment can be grouped into two time bins: nine events during the
first two seconds, then three more events after a time gap of six seconds.
Intriguingly, such a time structure is compatible with the modulation
of the neutrino signal by SN shocks, if the neutrino luminosities are
decreasing slowly with time. Thus, one might speculate that this
gap is connected with the passage of SN shocks through the
H-resonance. Then, the SN~1987A signal would be a hint that case B is
the true neutrino mixing scenario, i.e.\ that the neutrino mass
hierarchy is inverted.

\section*{Acknowledgments} 

We are grateful to Sergio Pastor for discussions that initiated the
Blackbox project. Supercomputer time at the John von Neumann Institute in 
J\"ulich is acknowledged. This work was supported by the European Science
Foundation (ESF) under the Network Grant No.~86 Neutrino Astrophysics,
and, in Munich, by the Deutsche Forschungsgemeinschaft (DFG)
under grant No.~SFB-375.  MK acknowledges an Emmy Noether fellowship
of the DFG, RT a Marie Curie fellowship of the European Union, and 
AD support by the MPI f\"ur Physik during a visit.

\section*{References}

\end{document}